\documentclass[journal=jacsat,manuscript=article]{achemso}

\usepackage[version=3]{mhchem} 

\author{Anastasiia Tukmakova}
\affiliation{TERAHERTZ PHOTONICS LLC, St. Petersburg, 191167, Russia}
\email{anastasiia_tukmakova@outlook.com}
\author{Patrizio Graziosi}
\affiliation{CNR - Institute for Nanostructured Materials (ISMN), Bologna, Italy}

\title[An \textsf{achemso} demo]
  {Machine learning unveils the materials physical properties driving thermoelectric generators efficiency: half-Heuslers case}

\abbreviations{TEG -- thermoelectric generator, TC -- thermocouple, TE -- thermoelectric, ML -- machine learning, DFT -- density functional theory, BTE -- Boltzmann transport equation, FEM -- finite elements method, HH -- half-Heusler alloys, TDF -- transport distribution function, density of states -- DOS, RAM -- random-access memory, CPU -- central processing unit, GPU -- graphics processing unit, GA -- genetic algorithm, CPM -- constant property model, PF -- power factor, avr. -- average}

\begin{document}



\begin{abstract}
We report the machine learning (ML)--based approach allowing thermoelectric generator (TEG) efficiency evaluation directly from 5 parameters: 2 physical properties -- carriers density and energy gap, and 3 engineering parameters -- external load resistance, TEG hot side temperature and leg height. Then, we propose to use genetic algorithm to optimize the proposed parameters in a way to maximize TEG efficiency.
To prepare data, physical properties of \textit{n}- and \textit{p}-type materials were computed by coupling Density Functional Theory to Boltzmann Transport, and used for Finite Elements simulations. TEG efficiency was evaluated from a finite elements model considering design, radiative heat loss, contacts, external load resistance and combinations of \textit{n}- and \textit{p}-type materials, as a result, giving 5300 different scenarios with the corresponding efficiency values.
For ML model, physical properties and engineering parameters were used as input features, skipping transport coefficients, while TEG efficiency was a target. 
Model was built on gradient boosting algorithm, its performance was evaluated using the coefficient of determination that reached a value of 0.98 on test dataset.
Features importance analysis revealed the most crucial features for Half-Heusler-based TEG efficiency: carriers density or Fermi level, indicating the predominant role of electrical conductivity and electronic part of electrical conductivity. Features that were less important, but able to  increase model performance were: energy gap, lattice thermal conductivity, charge carrier relaxation time and carriers conductivity effective mass. Features showed no impact were: density of states effective mass, heat capacity, density, relative permittivity and leg width.
The proposed approach can be applied for the identification of the most important physical properties and their optimal values, the optimization of TEG design and operation conditions in a way to maximize TEG efficiency.
\end{abstract}

\section{Introduction}
Solid-state thermoelectric generators (TEG) convert heat flow into useful electricity and can provide large benefits in the reduction of energy demand. Indeed, in line with some estimates, about 2/3 of all consumed energy is released as waste heat to the environment \cite{FORMAN20161568}. The harvesting of energy part with TEG technology can reduce our energy needs as well as a dependency on fossil fuels. Despite TEGs promising applications, their use is limited to niche applications for reasons of low efficiency and high prices. 

A TEG consists of an array of thermocouples (TC); a TC is a pair of \textit{p}- and \textit{n}-type legs of thermoelectric (TE) materials. The TEG output power (\textit{P}) and efficiency ($\eta$) depend on the operating conditions, the TEG design but, mostly, on the properties of TE materials. The latter are usually described in terms of a figure of merit, \textit{zT}, which is the function of temperature \(T\) and three temperature dependent material TE coefficients: electrical conductivity ($\sigma$), Seebeck coefficient (\textit{S}), and thermal conductivity ($\kappa$):
\begin{equation}
    zT = \frac{S^{2}\sigma}{\kappa}T\label{eqn:zT}
\end{equation}
These coefficients are intermediate phenomenological parameters, that are convenient for experimental materials characterization. The obtained \(zT\) values are usually used for a robust TEG maximum efficiency calculation in the framework of constant property model (CPM), which neglects temperature dependence of transport coefficients simplifying the estimation. Moreover, it does not consider operation conditions such as thermal loss due to the radiation or contacts impact. The equation \ref{eqn:eta_CPM} presents the formula for $\eta_{CPM}$ evaluation\cite{rowe}, where $Z_C$ is the figure of merit of a thermocouple, $T_h$ and $T_c$  are temperatures on the hot and cold sides of TEG,  $\overline{T}$ and $\Delta T$ are average temperature and temperature difference between $T_h$ and $T_c$.
\begin{equation}
    \eta_{CPM} = \frac{\Delta T}{T_h}\frac{\sqrt{1+Z_{C}\overline{T}}-1}{\sqrt{1+Z_{C}\overline{T}}+\frac{T_{c}}{T_{h}}}\label{eqn:eta_CPM}
\end{equation}

A promising shift can be done in the direction of TEG efficiency evaluation if temperature dependencies are taken into the account and transport coefficients are replaced by physical properties. These properties, e.g. carriers density, effective mass, energy gap, etc, have physical meaning unlike transport coefficients. The development of such a model could help the experimental research to find the specific properties to which an attention should be paid and estimate their desirable values.

From engineering point of view, the other encouraging improvement can be done if TEG design and operation conditions are considered within $\eta$ evaluation – heat loss, impact of contacts, active resistive load, legs design, legs segmentation, etc. These macro parameters form the temperature and electric potential distribution in the volume of TEG influencing the final thermoelectric effect.

A model is required that could reconcile all the aforementioned demands. However, the  problem is that such theoretical model estimating $\eta$ using physical properties as parameters still inevitably should consider transport phenomena in the material structure. This a theoretical description must be extremely sophisticated. Moreover, it would require an enormous amount of time and computing power to carry out such a simulation.

Machine learning (ML) being a data-driven approach can help to solve this problem by a direct connection between physical properties with TEG efficiency, skipping the intermediate simulations. ML does not require a theoretical description of all physical phenomena, it requires a significant amount of reliable and accurately prepared data being the key point of the model performance\cite{DL_phys_research}. Supervised machine learning methods are able to find hidden abstract patterns in labelled data and make predictions based on these patterns on new previously unseen data.

Literature analysis shows that, currently, literature dedicated to ML in TE field predictably tends to focus more on the search of materials with the best properties out of TEG context \cite{critical_rev, reg_guided_discvr, ML_for_mat_search}. The works in this field are dedicated to the search of physical properties from atomic scale descriptors and transport coefficients prediction.
For example, with the help of ML the optimal compositions of Cu-doped \ce{Bi-Te-Se} \cite{Bi_Te_Se_opt} has been found.  \ce{Fe-Pt-Sm}--based composition has been discovered as a material with a high Seebeck coefficient for a spin--driven thermoelectric effect \cite{spin_driven}.  Sasaki et.al\cite{Sasaki} proposed a model predicting optimal internal stress correlating with maximum Seebeck coefficient based on the analysis of thermal probe microimaging and micropoint X–ray diffraction scanning.  Furmanchuk et. al.\cite{Furmanchuk} developed a regression-based tool (ThermoEl toolkit) for Seebeck coefficient prediction in the range from 300 to 1000 K. ML was utilized for Half Heuslers band structure engineering\cite{HH_bands}. ML predicted\cite{feat_select} thermoelectric materials formula, found the most common elements in them and  predicted the most promising compositions from the selected elements. ML was proposed\cite{BiTe_similar} for a search of new materials with rhombohedral structure similar to \ce{Bi2Te3} and for the inverse design of doping in the \ce{(Bi$,$Sb)_2Te_3} family.\cite{ML_inverse_design}

Fewer papers address the questions of ML-based TEG optimization mainly focusing on design and using macroscopic parameters as input descriptors. In major part of works output data (values of efficiency or power) has been generated by finite elements method (FEM). FEM is a well-established and convenient tool acting as an alternative to experimental TEGs tailoring and test. FEM allows a simulation of TEG efficiency considering operation conditions, boundary conditions, losses and contacts. 
\ce{Bi2Te_{2.7}Se_{0.3}$-$ Bi_{0.5}Sb_{1.5}Te3}-based thermocouple geometry was optimized using deep learning\cite{ZHU2022117800} . 
TEG optimization has been implemented considering non-constant distribution of heat flux, losses at the contacts, and non-uniform current density\cite{ZAFAR2022131591}. 
 The optimization of legs area in the frustum leg thermoelectric generator has been implemented using ML and numerical simulations to increase its power and efficiency, at the same time reducing thermal stress\cite{ALOBAID2023119706}. 
ML allowed the optimization of TEG legs consisted of 7 segments of different composition, the optimal sequence was not in the strict correlation with ZT values of segments materials\cite{ZHAO2021114754}. 
Neural network, coupling genetic algorithm and FEM simulation were used to optimise leg height and width, TC fill factor and interconnect height considering different operating and contact resistance conditions\cite{ZHU2022117800}. 
A neural network has been utilized for a segmented leg to optimize the length and composition of low-, middle- and high-temperature layers\cite{DEMEKE20226633}. Authors concluded that for high temperature segments power factor (PF) is more crucial than zT. It was noteworthy the optimization of both power and efficiency, whereas normally indication on predominant importance of PF is mentioned in the context of power output problem\cite{Narducci_zT_vs_PF}.

Here, we propose a methodology for the direct prediction of TC efficiency based on materials physical properties. To prepare the data for ML study we carried out atomistic Density Functional Theory (DFT) simulations to obtain physical properties. The results of  DFT calculation were used as inputs to solve the Boltzmann Transport Equation (BTE) in the energy-, momentum-, and band--dependent relaxation time approximation. Thus, the temperature dependent transport coefficients were obtained, which next were used for a Finite Elements Method (FEM) simulation of TC efficiency, considering design, radiative loss, presence of contacts and external resistive load. The ML model was learning from materials physical properties and engineering TEG parameters, this data was matched with the TC efficiency, skipping the transport coefficients data.

\section{Methodology}
\begin{figure}
    \centering
    \includegraphics[width=0.75\linewidth]{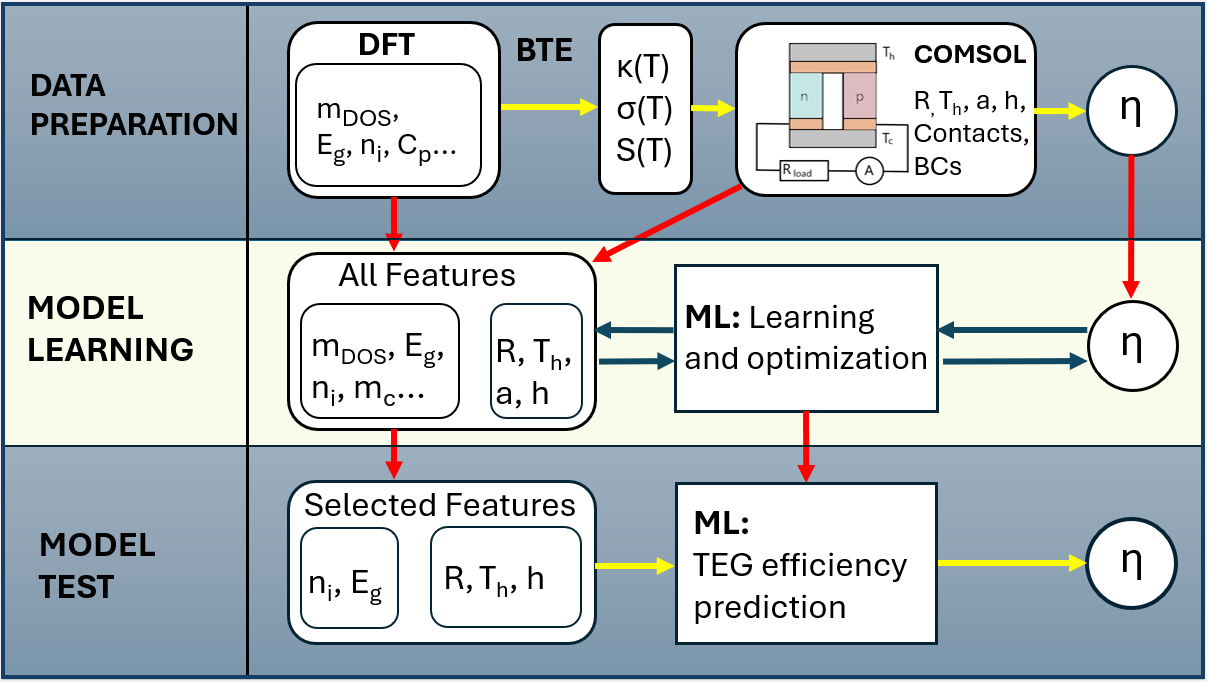}
    \caption{Research concept}
    \label{fig:concept}
\end{figure}

The research concept is presented in scheme (Figure \ref{fig:concept}). All data used for the simulations and ML models can be found in repository\cite{git_hub_rep_MLTEG}. The brief methodology is the following: 
\begin{enumerate}
    \item DFT simulations carrying out to calculate the electronic structure properties of materials
    \item The usage of DFT results as input for BTE to obtain temperature dependent transport coefficients: \(S, \sigma, \kappa\)
    \item The utilization of transport coefficients for numerical simulations to obtain TC efficiency considering engineering parameters 
    \item The ML model creation. The input data was physical properties and engineering parameters, the output -- TC efficiency. 
    \item Model development and optimization: feature engineering and selection, determination of required samples number, hyperparameters tuning. Model testing.
\end{enumerate}

\subsection{Materials and input dataset preparation}

\subsubsection{ Transport coefficients of TE materials from first-principles}
We considered a group of 14 half-Heusler alloys (HH), namely HfCoSb, HfNiSn, NbCoSn, NbFeSb, ScNiBi, ScNiSb, TiCoSb, TiNiSn, YNiBi, YNiSb, ZrCoBi, ZrCoSb, ZrNiPb, ZrNiSn. For each of them, we used the band structure computed with DFT and scattering parameters as in our previous reports,\cite{ACS2020} then we used the charge transport simulator ElecTra to solve the BTE and compute the charge transport coefficients electrical conductivity, Seebeck coefficient, and electronic contribution to the thermal conductivity.\cite{ElecTra, ELECTRA_GitHub} Importantly, ElecTra considers the full dependence of the relaxation time upon carrier momentum, energy and band index, as well as anisotropy in the scattering processes.\cite{JAP2019, ElecTra} In order to obtain a homogeneous set of data, all the material parameters have been computed under the same scattering considerations. Thus, we considered scattering with acoustic and non-polar optical phonons, the former is elastic, the latter inelastic, and scattering with ionised impurity dopants. All the scattering mechanisms have been considered to be both intra- and inter-band. Moreover, we adopted a full bipolar scheme.\cite{JPCC2020} In this way, we computed the charge transport coefficients for our HHs shortlist for several doping densities and in the temperature range from 300 K to 900 K. 

The solution of the BTE under the momentum relaxation time approximation, leads to the following expression for the charge transport coefficients:\cite{Phytos2011, Phytos2011b}
\begin{equation}
  \sigma_{ij(E_F,T)} = q_0^2 \int\limits_E \Xi_{ij(E,E_F,T)} \left( -\frac{\partial f_0}{\partial E} \right)dE  \label{eqn:sigma}
\end{equation} 
\begin{equation}
  S_{ij(E_F,T)} = \frac{q_0k_B}{\sigma_{ij(E_F,T)}} \int\limits_E \Xi_{ij(E,E_F,T)} \left( -\frac{\partial f_0}{\partial E} \right) \frac{E-E_F}{k_BT}dE  \label{eqn:S}
\end{equation}
\begin{equation}
  \kappa_{e,ij(E_F,T)} = \frac{1}{T} \int\limits_E \Xi_{ij(E,E_F,T)} \left( -\frac{\partial f_0}{\partial E} \right) \left(E-E_F\right)^2dE - \sigma_{ij(E_F,T)}S^2T  \label{eqn:kappa}
\end{equation}
where $\Xi_{ij(E,E_F,T)}$ is the Transport Distribution Function (TDF) defined below in equation \ref{eqn:TDF}, $E_F$, $T$, $q_0$, $k_B$, and $f_0$, are the Fermi level, absolute temperature, electronic charge, Boltzmann constant, and equilibrium Fermi distribution, respectively.
The TDF is computed as
\begin{equation}
  \Xi_{ij(E,E_F,T)} = \frac{2}{(2\pi)^3} \sum_n\sum_{\textbf{\textit{k}}_{nE}} v_{i,\textbf{\textit{k}}_{nE}}v_{j,\textbf{\textit{k}}_{nE}}\tau_{i,\textbf{\textit{k}}_{nE}} g_{i,\textbf{\textit{k}}_{nE}}  \label{eqn:TDF}
\end{equation}
In equation \ref{eqn:TDF} $v$ is the band velocity, $\tau$ the relaxation time, and $g$ the electronic desnity of states (DOS). All these quantities are specific of each individual transport state $\textbf{\textit{k}}_{nE}$, where $\textbf{\textit{k}}$ is the wave-vector, $n$ indicates the band index and $E$ its energy. So, the sum runs over all the transport states identified by their momentum, belonging to all the bands, having a certain energy. The equations that define these quantities in details can be found in previous publications.\cite{JAP2019,ElecTra}

For the lattice contribution to thermal conductivity, we based on available experimental or computational data at 300 K \cite{kappa_HfCoSb_TiCoSb_ZrCoSb, kappa_HfNiSn_TiNiSn, kappa_NbCoSn, kappa_NbFeSb_TiCoSb_ZrCoBi_ZrCoSb_ZrNiPb_ZrNiSn, kappa_ScNiBi, kappa_ScNiSb, kappa_YNiSb, kappa_YNiSb_2} and assumed a $\frac{1}{T}$ scaling,\cite{ACS2024} and then we summed the electronic contribution under the assumption that only the latter depends upon carrier density.
For the heat capacity, we used available data in literature.\cite{Ch_HfCoSb_TiCoSb_ZrCoSb, Ch_NbCoSn, Ch_NbFeSb,Ch_ScNiSb,Ch_YNiBi} In case of absence of literature data, we adopted the Dulong-Petit limit of 74.8 J/(K mol),\cite{Osafile_2020} for 300 K and 900 K, and a linear variation between these two extrema. The mass density is kept from the experimental literature and, whenever absent, has been computed from the mass and volume of the unit cell. For this reason, and for consistency with the scattering rates calculations, our previous studies, and amongst the materials for lack of data, we neglected the temperature dependence of both the mass density and the dielectric constant.

Thus, we have the necessary temperature dependent parameters to be passed to COMSOL Multiphysics {\textregistered} (COMSOL). Because these coefficients depend also on the carrier density $n_i$, we chose three reference values, $5\times10^{18}$, $5\times10^{19}$, and $5\times10^{20}$ cm$^{-3}$, and for each material we selected the closer computed carrier concentration. These values have been selected because they are in the range of the experimental values and are in the available range of the considered materials. Because we did this for both n-type and p-type conditions, we obtained 42 \textit{n-}type materials and 42 \textit{p-}type materials for a total of 1764 possible combinations.

About the material features to be used in the subsequent ML studies, a part of them has been taken from the data directly used by COMSOL, others have been obtained analysing the electronic structure or the ElecTra code output. Amongst the former, we have the band gap and the effective masses, obtained with the EMAF code,\cite{EMAF_GitHub} amongst the latter, we have the carrier density and the effective relaxation time $\tau_{eff}$, which is evaluated as 
\begin{equation}
  \tau_{eff(n_i,T)} = \frac{\int\limits_E \tau_{x(E,n_i,T)}g_{(E)}dE}{\int\limits_Eg_{(E)}dE} \label{eqn:tau_eff} 
\end{equation}
with $\tau_{x/y/z}$ being identical for the cubic symmetry of the considered HH.
When a certain feature $f$ was depending on temperature, for the ML studies we averaged it as:
\begin{equation}
  f_{ave} = \frac{\int\limits_T f_{(T)}dT}{\Delta T} \label{eqn:T_ave} .
\end{equation}
Such an average has been carried on two reasonable temperature ranges, 300 to 600 K and 300 to 900 K. This is to consider the temperature dependence of the feature and at the same time to use features which are general of a material.

\subsection{Numerical simulation and output data generation}
\subsubsection{Geometry and Materials}
\begin{figure}[H]
\begin{minipage}[l]{0.34\linewidth}
\includegraphics[width=1\linewidth]{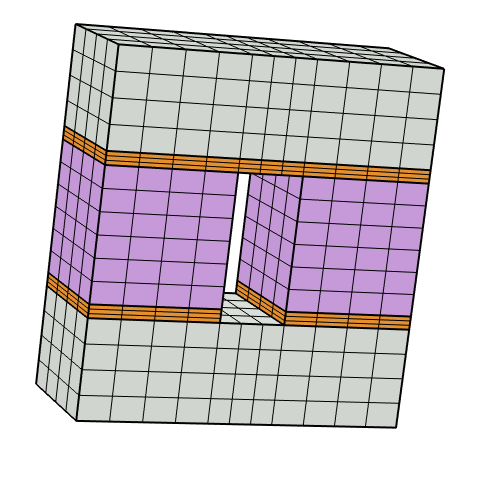} \\ \textbf{a} 
\end{minipage}
\hfill
\begin{minipage}[l]{0.31\linewidth}
\includegraphics[width=1\linewidth]{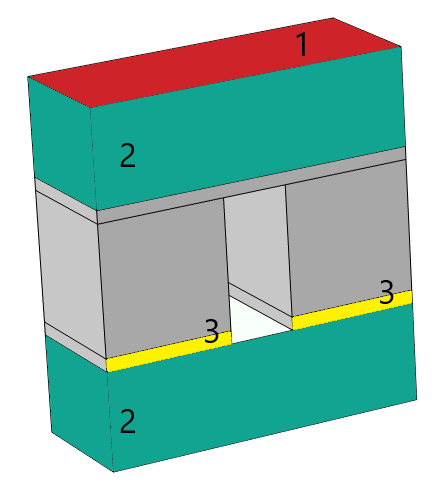} \\ \textbf{b} 
\end{minipage}
\hfill
\begin{minipage}[l]{0.31\linewidth}
\includegraphics[width=1\linewidth]{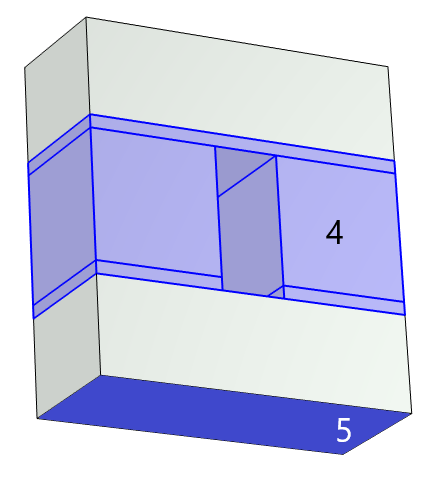} \\ \textbf{c} 
\end{minipage}
\caption{a) Thermocouple model with the built mesh; grey -- ceramic plates, purple -- TE legs, orange -- metal contacts. Boundary conditions: b) 1 (red) -- constant temperature $T_{h}$, 2 (green) -- specified temperature, 3 (yellow) -- the conditions for external electrical circuit: ground and terminal c) 4 (light blue) -- radiative heat flux, 5 (deep blue) -- constant temperature $T_{c}=300 K$}
\label{fig:mesh_BCs}
\end{figure}

The simulated object (Figure \ref{fig:mesh_BCs} a) was 1 thermocouple consisted of 2 ceramic \ce{Al2O3} plates, metal \ce{Cu} contacts and 2 thermoelectric legs of \textit{p-} and \textit{n-}type. Simulation of a thermocouple is enough for efficiency evaluation as efficiencies of one TC and a whole TEG are the same. Legs materials, height and width were varying within the simulation.  Legs width was equal to the length and was changing in the range from 1 to 5 mm; legs height was changing from 1 to 10 mm.
Ceramic plates had a fixed height of 1.5 mm, metal contacts had a fixed height of 0.2 mm. Their width and length were altered respectively to the legs cross section change.
Materials were described via thermal conductivity, electrical conductivity, Seebeck coefficient, heat capacity, mass density, relative permittivity. Properties of \ce{Al2O3} and \ce{Cu} were taken from the built-in library. Data about TE materials used for simulation can be found here. All properties were considered to be functions of temperature. 

\subsubsection{Mesh}
\begin{figure}[H]
    \centering
    \includegraphics[width=0.75\linewidth]{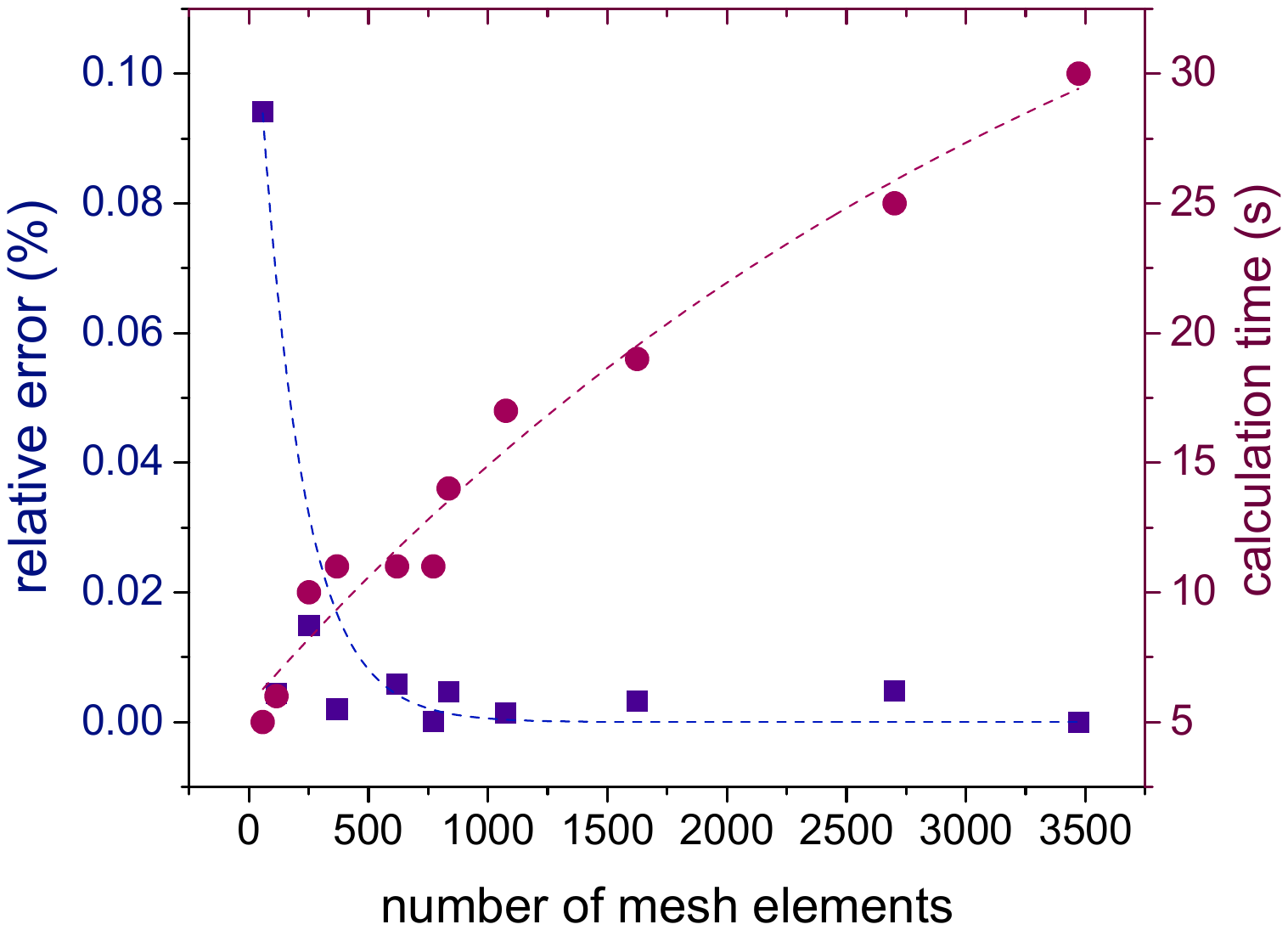}
    \caption{The dependencies of relative error of temperature evaluation and time of 1 model run on number of mesh elements.}
    \label{fig:num_el}
\end{figure}

Mesh was manually controlled, based on a mapped layer and swept through the whole volume of the object. The number of elements has been selected based on their impact on relative error  (Figure \ref{fig:num_el}). Relative error was calculated as \( \left|
\frac{\Delta T_{leg\;ref} - \Delta T_{leg}}{\Delta T_{leg\;ref}} \cdot 100 \% \right| \). $\Delta T$ is a temperature difference in TE legs (average value between \textit{n-} and \textit{p-}type legs). $\Delta T_{leg\;ref}$ -- is the $\Delta T$ calculated for a model with the largest number of mesh elements (3472) taken as a reference.

We selected a reasonable balance between calculation error and speed. The object was split on 772 mesh elements. There was a fixed number of mesh elements along z-axis: 4 for ceramic plates, 3 for metal contacts and 6 for thermoelectric legs (Figure \ref{fig:mesh_BCs} a) .  In the x-y plane the size of the mesh elements have been adjusted and set to be of a “normal” size. 

Further refinement of mesh would not lead to a significant relative error decrease, but would result in calculation time prolongation. At the same time the chosen mesh allowed faster simulation with a time for 1 run of approximately 11 seconds. The calculations were carried out on a laptop with 12 Gb RAM, equipped with Intel Core i7-8565U CPU (1.80 GHz, 4 cores, 8 threads), NVIDIA GeForce MX150 GPU (2 GB VRAM) and Intel UHD Graphics 620 (1 GB VRAM). The model with the described setup spends 16--17 hours on all 5300 runs.

\subsubsection{Mathematical Model}

For the model creation we used 3 COMSOL physical interfaces -- "Heat transfer in solids", "Electric Current", "Electrical Circuit" and 1 multiphysical interface -- "Thermoelectricity" . The dependent variables of the task were temperature and electric potential fields in the TC and current in the electrical circuit. 
System of differential equations is presented in Eq. \ref{eqn:en_bal}  --\ref{eqn:q}.
Energy and charge conservation laws:
\begin{equation}
    \mathbf{\nabla} \cdot \mathbf{q} - Q_{j} = 0\label{eqn:en_bal}
\end{equation}
\begin{equation}
    \mathbf{\nabla} \cdot \mathbf{j} = 0\label{eqn:charge_bal}
\end{equation}
Current and heat flux:
\begin{equation}
  \mathbf{j}=\sigma (\nabla V + S \nabla T) \label{eqn:j}
\end{equation}
\begin{equation}
  \mathbf{q}=-\kappa \nabla T + S T\mathbf{j} \label{eqn:q}
\end{equation}
where \(\mathbf{j}\) -- current density, \(\mathbf{q}\) -- heat flux,  \(Q_{j}\) -- heat source caused by Joule heating, \(T\)-- absolute temperature, \(V\) -- electric potential.

    Initial values of temperature and electric voltage were: \(T_{init} = 293\;K\), \(V_{init} = 0\;V\).

    Thermal boundary conditions included radiative heat flux, specified temperature and constant temperature values on hot and cold sides on a thermocouple. Specified temperature was equal to \(T\). We did not include convective heat flux on the lateral surfaces as TEGs are encapsulated excluding the possibility of convection. 
    The radiative heat flux:
\begin{equation}
  -\mathbf{n} \cdot \mathbf{q}=\varepsilon_{em} \sigma_{SB} (T_{amb}^4 - T^4) \label{eqn:q_rad}
\end{equation}   
where $\sigma_{SB}$ is a Stephan-Boltzmann constant, $T_{amb}$ -- ambient temperature, $\varepsilon_{em}$ is a surface emissivity (0.82 for TE legs, 0.8 for \ce{Al2O3} and 0.5 for \ce{Cu}).

    Electric boundary conditions included electrical insulation of the whole object except for two contacts that were connected to the external electrical circuit. For this purpose terminal boundary condition (type "circuit") was used. The circuit consisted of ground node, resistor, an ampere meter to track the current, and external terminal. The derived electric potential from electric currents interface was taken as a terminal voltage. 
    
    To calculate the TEG efficiency we used the following formula\cite{rowe} : \(\eta = \frac{P}{Q}100\%\), where \(P = I^2 R\) with \(R\) being a defined parameter and \(I\) calculated by the model. The value of \(Q\) was evaluated as the normal total heat flux (W) on the hot side of the TC at the post processing stage.
    
\subsection{Machine Learning}

\subsubsection{Data}
Parameters in the input data fed to ML model are usually called “features”. The role of features were played by data obtained from DFT along with engineering parameters: 
    \begin{itemize}
        \item carriers density, $n_i$, 
        \item Fermi level, $\eta_F$
        \item energy gap, $E_g$
        \item density of states effective mass, $m_{DOS}$
        \item minimal density of states effective mass, $m_{DOS\;min}$
        \item carriers conductivity effective mass, $m_c$
        \item charge carrier relaxation time, $\tau_{eff}$
        \item density, $\rho$
        \item heat capacity at constant pressure, $C_p$
        \item relative permittivity, $\varepsilon$
        \item temperature on the hot side of TC, $T_h$
        \item resistance on the external load, \(R\)
        \item leg height, \(h\)
        \item leg width, \(a\)
    \end{itemize}
    Transport coefficients were not used as features at this stage. ML is supposed to be able to find direct connection between physical and engineering properties and TC efficiency, skipping the intermediate calculations. 
    
\begin{figure}[H]
  \includegraphics[width=0.8\linewidth]{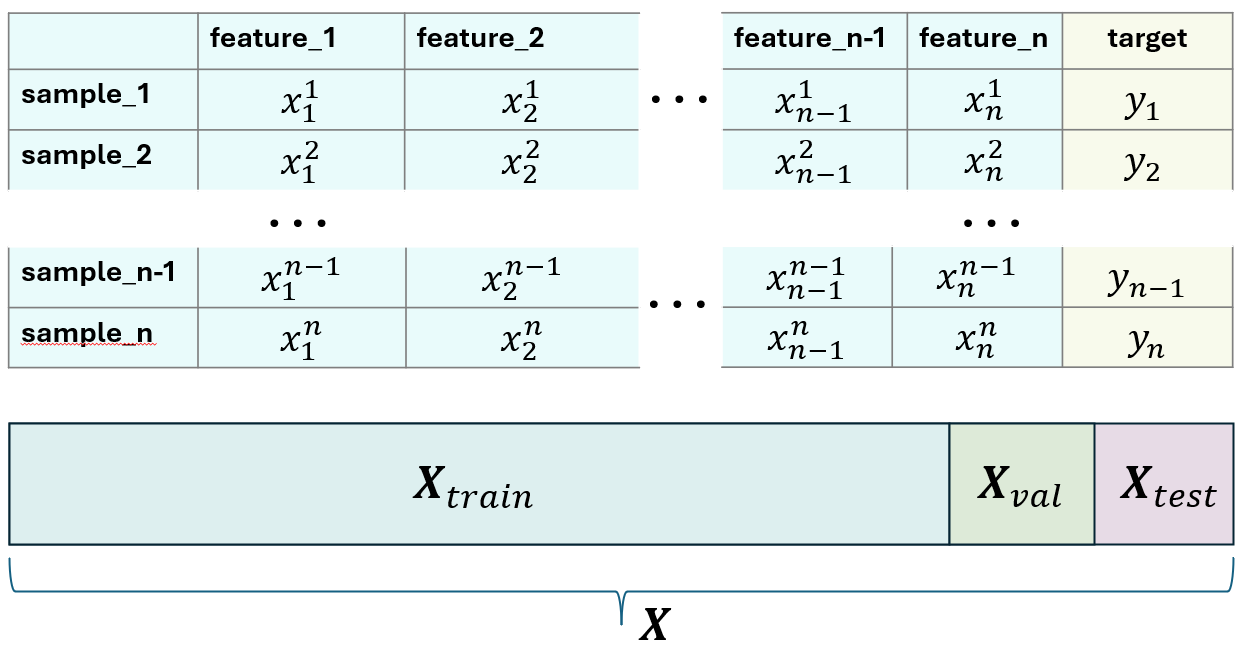} \\ 
  \caption{Data structure representation}
  \label{fgr:data_structure}
\end{figure}

     In classical ML, features form the table (Figure \ref{fgr:data_structure}) that can be also called a feature matrix and usually denoted as \textbf{X}. In such representation of data, each column corresponds to a specific feature, while each row refers to a single sample. By "sample" we assume a single scenario of system behaviour. In our case, each sample is characterized by \textit{n}- and \textit{p}-type material properties, TEG design and operation conditions. In other words, each sample is a unique combination of features resulting in a specific TEG efficiency. The dimensionality of \textbf{X} matrix for $n$ features is $5300 \times n$.
 The \textbf{X} matrix is combined with a target vector denoted as \textbf{y}. In our task, target vector is formed from all efficiency values for every sample, hence, having a dimensionality $5300 \times 1$.

 The initial dataset \textbf{X} (5300 samples) was split into test, validation and training dataset: \textbf{X}$_{test}$, \textbf{X}$_{val}$, \textbf{X}$_{train}$(Figure \ref{fgr:data_structure}). \textbf{X}$_{test}$ accounted for 10\% of \textbf{X} (530 samples), \textbf{X}$_{val}$ for 10\% of \textbf{X}$_{train}$ + \textbf{X}$_{val}$ (477 samples). The rest data made up the biggest part of the whole being the training dataset from which the model learns the patterns between data (4293 samples). \textbf{X}$_{val}$ dataset is used to check the performance of the model learned from \textbf{X}$_{train}$, if metric on \textbf{X}$_{val}$ is worse than on \textbf{X}$_{train}$ that means that model is overfitted to the train data. In this case, parameters of the model are changed, it is fitted to the \textbf{X}$_{train}$ data again, and then again the performance is evaluated on validation data. After the model is optimized and metrics on both \textbf{X}$_{train}$ and \textbf{X}$_{val}$ are high and close to each other, model finally tested on \textbf{X}$_{test}$ dataset that model haven't seen before. If metric is high and close to the ones obtained within training and validation, it is concluded that models has reached high generalization ability and can show high performance on new data.

\subsubsection{Correlation between features}
    The analysis of correlations shows if some features have high or low linear correlation. In major cases, it is recommended to delete one of two features if they are highly correlated. High linear correlation reveals that features carry the same information, so they are assumed to be redundant. High linear correlations can lead to the problems with model convergence, performance decrease, calculation time prolongation. 
    
\begin{figure}[H]
    \centering
    \includegraphics[width=1\linewidth]{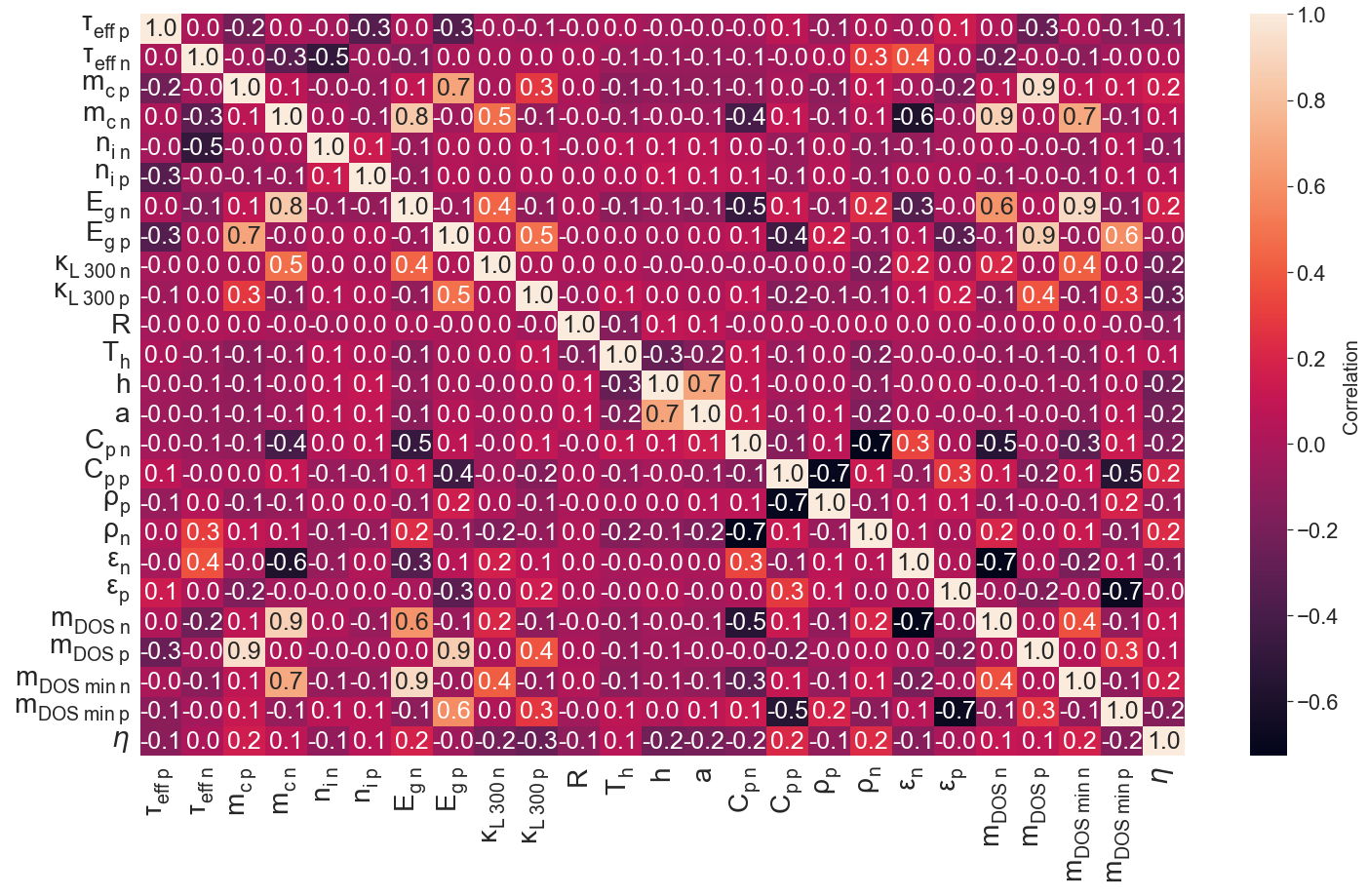}
    \caption{Correlation Matrix. Linear correlation coefficients between features.}
    \label{fig:corr_matr}
\end{figure}

    We built the correlation matrix of all considered features. It is presented in Figure \ref{fig:corr_matr}, correlation coefficient values have been rounded to 1 decimal place to fit the figure. The information about correlations were taking into the consideration within the feature importance analysis (Section "Feature selection"). For example,  $m_{DOS\;p}$ has very high correlation with $m_{c\;p}$ and $E_{g\;p}$ of about 0.9. That means that, more likely, $m_{DOS\;p}$ is redundant and can be deleted, end the same for $m_{DOS\;p}$. Indeed, we treated symmetrically \textit{n-} and \textit{p-}type features.
 
\subsubsection{Number of samples}
By the definition, machine learning is a program whose performance increases with experience. By experience we mean the amount of data that model “sees”. This is why it is so crucial to find the optimal quantity of data. The lack of samples results in a weak performance even with the strong algorithms used. On the other hand, some point exists beyond which the enlargement of samples number results in such insignificant performance improvement that further dataset gain does not lead to a noticeable practical benefit. 

\begin{figure}[H]
    \centering
    \includegraphics[width=0.75\linewidth]{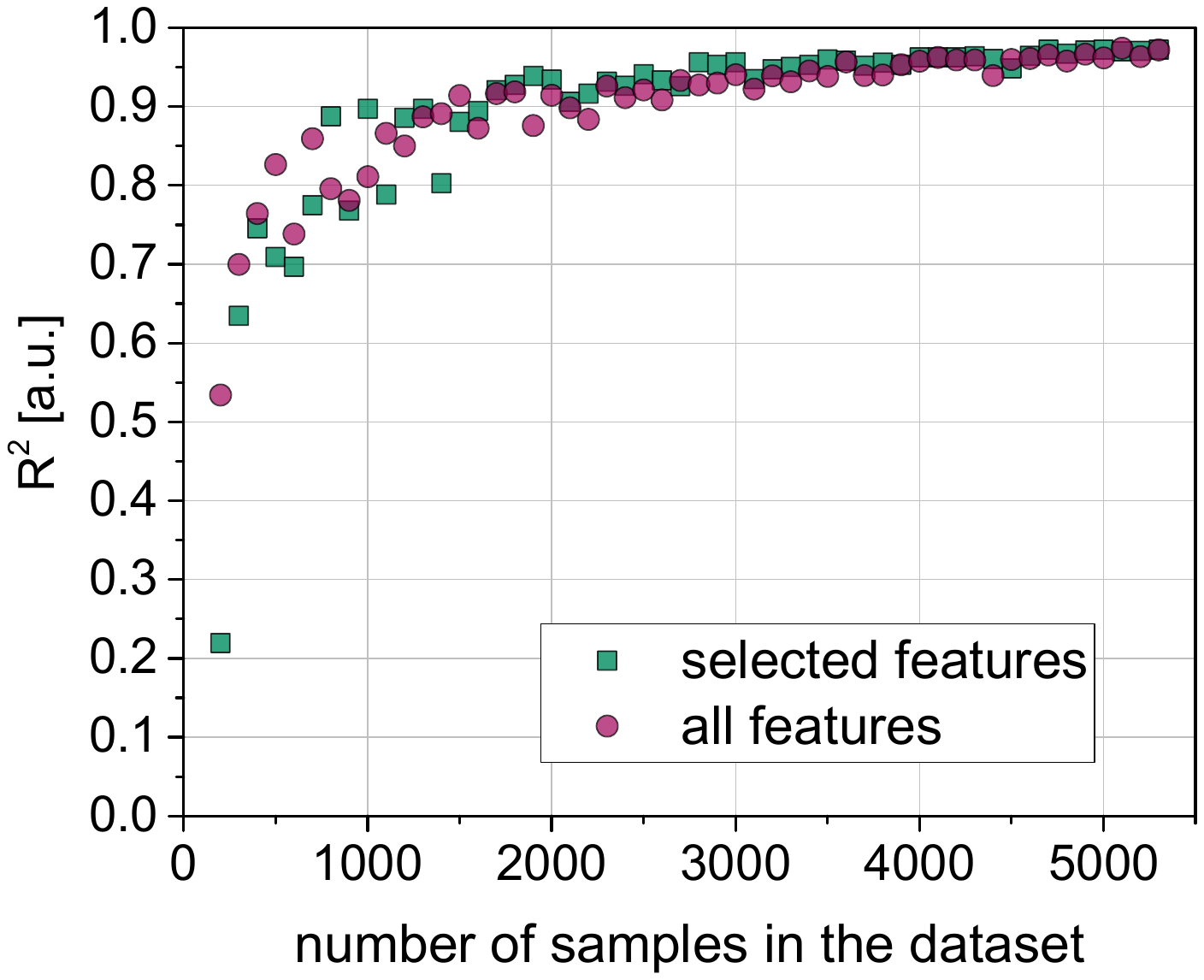}
    \caption{The dependence of \(R^2\) on dataset size}
    \label{fig:r2_num_samples}
\end{figure}

We tested the $R^2$ metric on test dataset using a model that was trained on different samples number, and present the dependence of $R^2$ on number of samples in the whole dataset \textbf{X} (Figure \ref{fig:r2_num_samples}). All features were used for this model. We were adding new data in little portions of several hundreds of samples as long as a change in metric was noticeable. In the range from 4000 to 5000 samples the growth of \(R^2\) became slow, around 1\% per 1000 samples. At the same time, in average, \(R^2\) value was constantly higher than 0.95. So, 5000 samples seem to be a reasonable quantity allowing to obtain \(R^2\) value of about 0.97 -- 0.98. We used 5300 samples for learning.

Then, we checked this pattern for the model based only on selected features.  After 3000 samples are reached, almost no difference between the models can be seen.

\subsubsection{Task formulation and algorithm selection}

    The purpose of the model is to predict TEG efficiency values based on the labelled dataset, that means we should use supervised machine learning for a regression problem. We chose decision trees algorithm\cite{sklearn_DT} for this task that is available in scikit-learn library written for Python. This algorithm is easy to work with, e.g. it allows the user to skip such data preprocessing stages as normalisation, standardisation, log-transformation or gaps filling. 
However, there are some disadvantages of a model based on one decision tree: it tends to overfit and to be a “weak learner”. This is why usually an ensemble of trees is used, helping to overcome both of the mentioned problems. 

Taking into the cosideration all mentioned above, we selected gradient boosting regressor, that is an ensemble of decision trees, as an algorithm for our task. This algorithm is also implemented in  scikit-learn learn library\cite{sklearn_GB}.
Gradient boosting  is a powerful tool that combines weak models sequentially in the way that every next model learns on the mistakes of the previous one. It also provides features importances analysis which is of great help for our task as it provides the opportunity for a convenient model interpretation.  

\subsubsection{Hyperparameters tuning}

The hyperparameters are the parameters of the model whose adjustment helps to improve the performance and prevent overfitting.
The hyperparameters were selected utilising the optuna\cite{optuna} library. Learning was based on training dataset, the optimization was based on the $R^2$ metric minimization on validation dataset. The selection of hyperparameters range showed to have an impact on final metric. The lowest and highest boundaries of hyperparameters values have been chosen manually by brute force empirically and presented in table. The number of trials was set to be 20.

\begin{table}
    \centering
    \begin{tabular}{ccc}
    \hline
         hyperparameter&  \multicolumn{2}{c}{range}\\
         &  from&  to\\
         \hline
         learning rate&  0.04
&  0.25
\\
         number of estimators&  350
&  600
\\
         maximum depth&  1
&  15
\\
         minimum samples leaf&  30
&  40
\\
         minimum samples split&  2
&  40
\\
    \hline
    \end{tabular}
    \caption{Hyperparameters and ranges of their values that have been used for model optimization.}
    \label{tab:hyperparams}
\end{table}

\subsection{Features selection}
    
    Features selection is a crucial part for ML models. For some ML methods, e.g. neural networks or logistic regression, the performance strongly correlates with the statistical distribution of data. In such models, the removal of excess features helps to offset the multicollinearity problem that can ruin the model efficiency in 2 ways: reducing the metrics and expanding calculation time. The utilisation of decision trees helps to overcome the first problem: high correlation between features does not crucially affect the metric. But redundant features complicate the model leading to overfitting and lower stability (weaker performance on unseen data) and longer calculation time. They can also decrease the interpretability of the model. Hence,  it is recommended to delete the redundant features. 
    However, there is another aspect of features importance analysis beyond practical sense. It allows a wide perspective for interpretation and insights. Features importance shows how strongly a particular feature impacts on the loss function minimization that describes how close are predicted values to the actual ones. So, important features help to build a better regression model and to describe a target function, while not important features do not have a contribution. In our particular case, features importance analysis shows a higher or lower impact of every particular feature on TEG efficiency. 

    We considered models based on Fermi level position ($\eta_F$) and carriers density ($n_i$), one instead of the other due to the high correlation between the twos that we observed and the fact that they carry nearly the same physical information, and models based on features avereged in the ranges from 300 to 600 K and from 300 to 900 K. The model estimation was based on $R^{2}$ metric obtained on data unseen by the model within the learning ($\textbf{X}_{test}$). 
The process of features selection consisted of several steps. First of all, model based on all features has been created and optimized via the hyperparameters adjustment. 

\begin{figure}[H]
\begin{minipage}[l]{0.49\linewidth}
\includegraphics[width=1\linewidth]{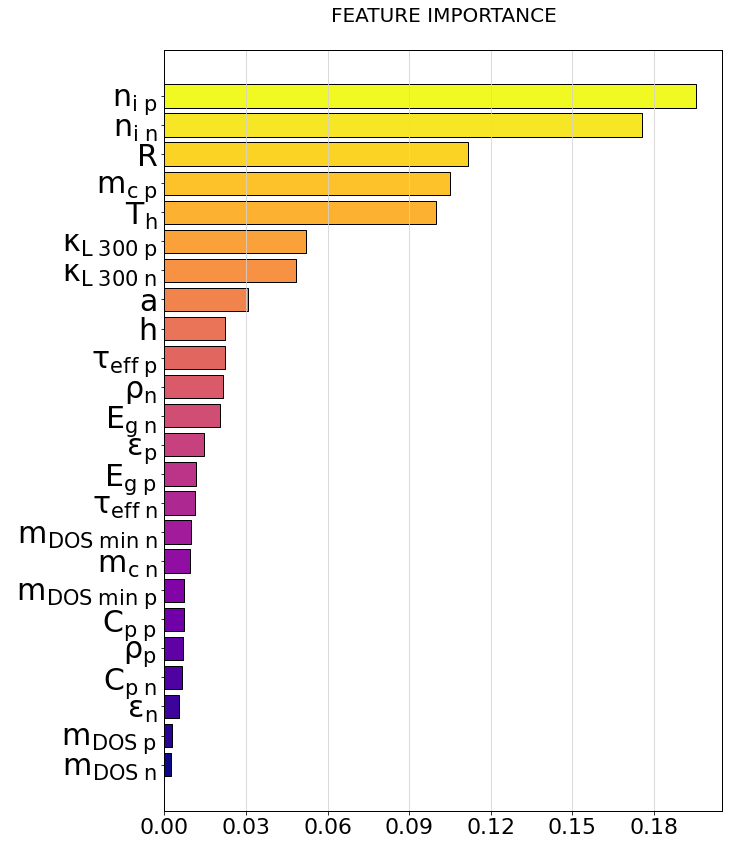} \\ \textbf{a} $T_{h\;avr} = 600 K$
\end{minipage}
\hfill
\begin{minipage}[l]{0.49\linewidth}
\includegraphics[width=1\linewidth]{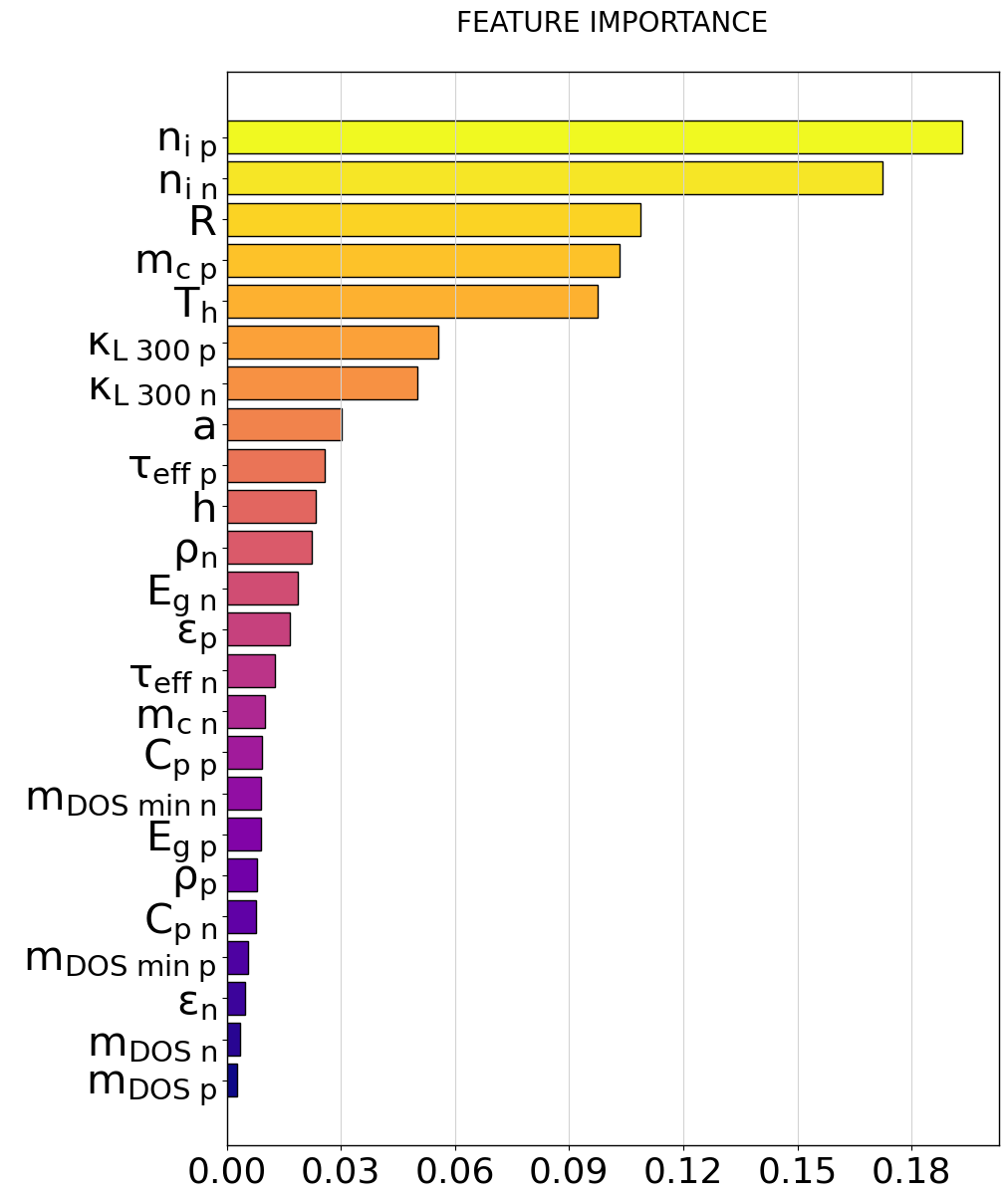} \\ \textbf{b} $T_{h\;avr} = 900 K$
\end{minipage}
\caption{Features importance for a model based on carriers density including all features}
\label{fig:feat_imp_n_i_all}
\end{figure}

\begin{figure}[H]
\begin{minipage}[l]{0.49\linewidth}
\includegraphics[width=1\linewidth]{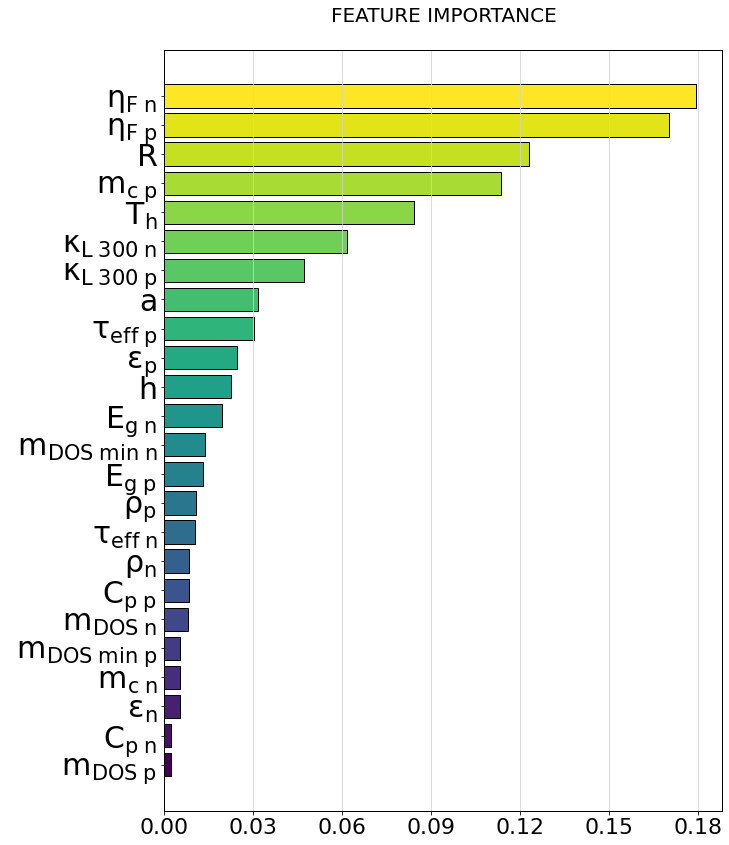} \\ \textbf{a} $T_{h\;avr} = 600 K$
\end{minipage}
\hfill
\begin{minipage}[l]{0.49\linewidth}
\includegraphics[width=1\linewidth]{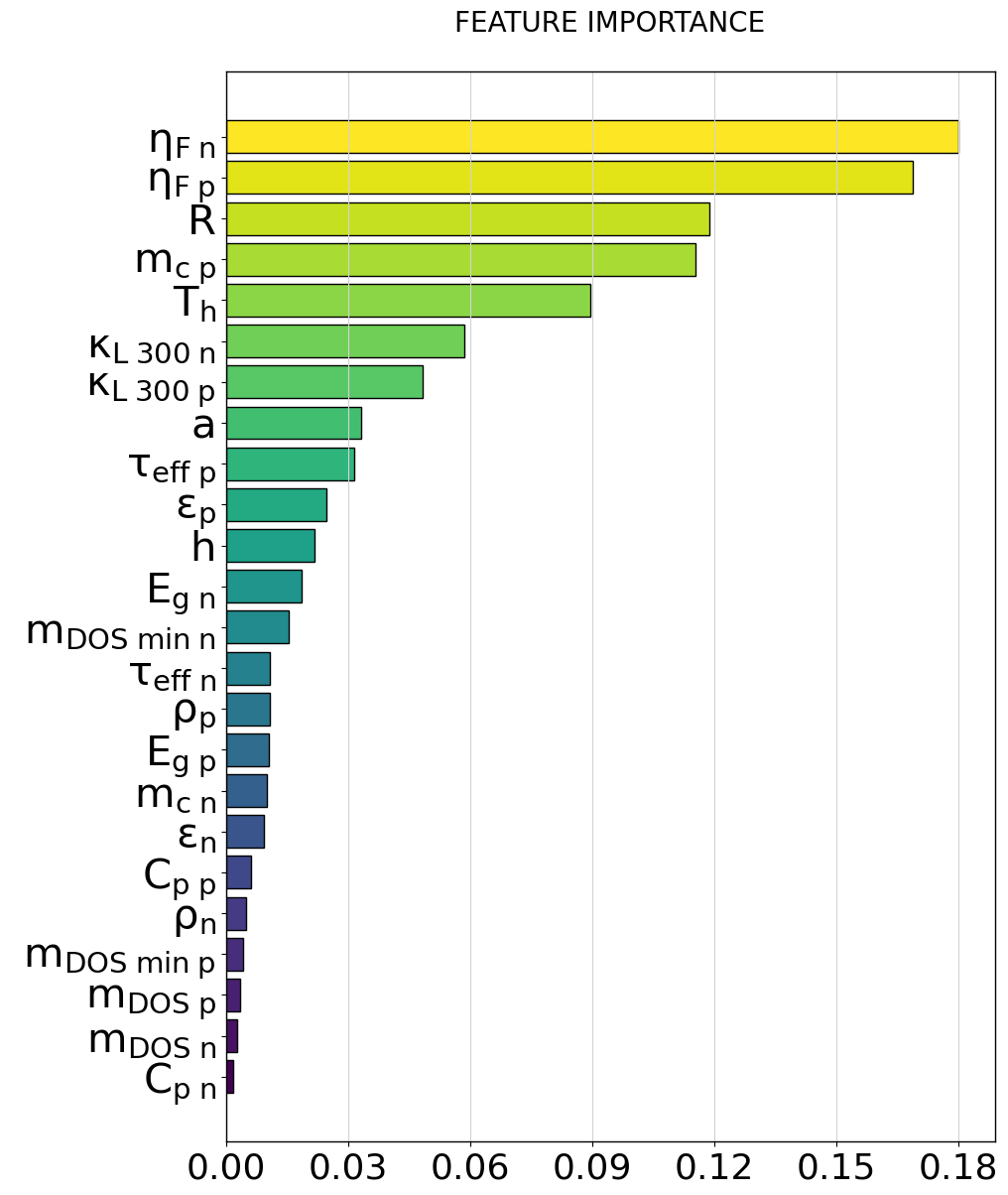} \\ \textbf{b} $T_{h\;avr} = 900 K$
\end{minipage}
\caption{Features importance for a model based on Fermi level including all features}
\label{fig:feat_imp_eta_F_all}
\end{figure}

\begin{figure}[H]
    \centering
    \includegraphics[width=0.5\linewidth]{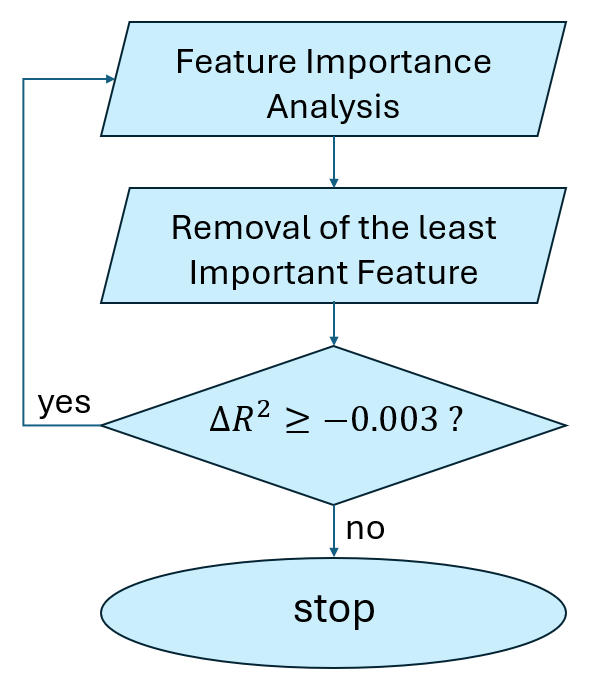}
    \caption{Feature selection algorithm}
    \label{fig:feat_sel_algo}
\end{figure}

Then, the feature importance analysis has been done (Figures \ref{fig:feat_imp_n_i_all} and \ref{fig:feat_imp_eta_F_all}). The values of importances show how each feature is important in relation to the others. For example, $n_{i\;p}$ has an importance of about 0.2, while $\varepsilon_p$ of about 0.02. That means, that for this specific model  $n_{i\;p}$ has 10 times higher impact on model performance than $\varepsilon_p$.
Sum of all importances gives 1.0.  
Within the feature selection process (Figure \ref{fig:feat_sel_algo}) that can be called "bottom-to-top", the least important feature is deleted. Then, feature importance analysis is done again to reveal the next least important feature. The removal of one feature can change the distribution of importances and it changes the model performance, this is why we did this process step-by-step. After each step we evaluated the change of metric $\Delta R^{2}$ on test dataset to check its values. We set the threshold of $\Delta R^{2}$ to be \(-0.003\) .  All features that have been deleted resulted in no metric change or caused an insignificant drop with $\Delta R^{2} \geq -0.0003$. These features will be addressed as the least important.
Additionally, we performed a "top-to-the-bottom" feature selection excluding the least important features revealed by the previous selection. Based on the previous analysis it became obvious that the most important features are carriers density and Fermi level. Two models using a single feature have been created: one based on  $n_i$ and the other based on $\eta_F$. The $R^2$ has been evaluated for these baseline models. Then, 7 models using different second feature were created and the changes in metric were tracked (Table \ref{tab:2_features}). This approach helped to evaluate the impact of every single feature and distinguish the most important and less important features. We assume the most important features (in addition to the $n_i$ and $\eta_F$ ) to be $T_h$, \textit{R} and \textit{h}.

 \begin{table}[H]
     \centering
     \begin{tabular}{cccc}
     \hline
 \multicolumn{2}{c}{1st feature -- $n_i$}& \multicolumn{2}{c}{1st feature -- $\eta_F$}\\
 \hline
          2nd feature&  \textbf{\(R^2\)} 
&  2nd feature& \textbf{\(R^2\) } 
\\
          \hline
          --&  0.63 
&  --& 0.49 
\\
          $\tau_{eff}$&  0.62 
&  $\tau_{eff}$& 0.62 
\\
          $m_c$&  0.63 
&  $E_g$& 0.62 
\\
          $E_g$&  0.64 
&  $\kappa_{L\;300}$& 0.63 
\\
          $\kappa_{L\;300}$&  0.64 
&  $m_c$& 0.63 
\\
          h 
&  0.82 
&  h 
& 0.66 
\\
          R 
&  0.88  
&  
R& 0.72 
\\
          $T_h$&  0.90  &  $T_h$& 0.75 \\
      \hline
     \end{tabular}
     \caption{The comparison of the performance of the models based on 2 features.}
     \label{tab:2_features}

 \end{table}
Then, we created a model based on 4 features -- 1 physical ($n_i$ or $\eta_F$) and 3 engineering ($T_h$, \textit{R} and \textit{h}). We added one more feature: $\tau_{eff}$, $m_c$, $\kappa_{L\;300}$ and $E_g$ to check which of them allows the best metric. The analysis (Table \ref{tab:5_feat}) revealed that energy gap helps to maximize $R^2$.

\begin{table}[H]
    \centering
    \begin{tabular}{cccccc}
    \hline
         basic features set&  2nd physical feature&  $R^2$&  basic features set&  2nd physical feature& $R^2$\\
         \hline
         &  – 
&  0.952 
&  &  – 
& 0.800 
\\
         &  $\tau_{eff}$&  0.958 
&  &  $\tau_{eff}$& 0.957 
\\
         $n_i$ ,  $T_h$, R, h&  $m_c$&  0.963 
&  $\eta_F$, $T_h$, R, h&  $\kappa_{L\;300}$& 0.973 
\\
         &  $\kappa_{L\;300}$&  0.972 
&  &  $m_c$& 0.976 
\\
         &  $E_g$&  0.979 &  &  $E_g$& 0.981 \\
    \hline
    \end{tabular}
    \caption{Performance of the models based on 5 features: 2 physical and 3 engineering.}
    \label{tab:5_feat}
\end{table}

The final result of features importance analysis can be seen in Table \ref{tab:feat_status}.
 \begin{table}[H]
    \centering
    \begin{tabular}{cc}
    \hline
 feature status&feature name\\
 \hline
         the most important& $n_i$ or $\eta_F$, $T_h$, R, h\\
         less important& $E_g$, $m_c$, $\kappa_{L\;300}$, $\tau_{eff}$\\
         the least important& $m_{DOS\;min}$, $m_{DOS}$, $\rho$, $C_p$, $\varepsilon$, a,\\
    \hline
    \end{tabular}
    \caption{features divided into 3 groups in accordance with their importance analysis}
    \label{tab:feat_status}
\end{table}

\section{Results and discussion}
\subsection{Model metrics analysis}
The obtained $R^{2}$ values for the models are presented in Table \ref{tab:r2_vals}. Metrics presented in the "Selected Features" column are given for a model built on 2 physical parameters -- $n_i$ or $\eta_F$ and $E_g$, and 3 engineering parameters -- $T_h$, R, h. Models built on only selected features show not worse, but even better performance than model containing all features. The $R^{2}$  improvement is  \(0.5 - 0.7 \%\). 
Table contains data about models using physical features averaged over 300 -- 600 K or 300 -- 900 K. It is worth noting that this factor have a negligible impact on metrics, as no significant difference in $R^2$ can be seen. Hence, properties for the model can be averaged in either of two temperature ranges.

Within the process of results analysis we observed that small variations of $R^{2}$ of about \(0.1\; \% \) are possible; this can be explained by the optuna optimizer usage that utilizes the random initialization of hyperparameters.
\begin{table}

    \centering
    \begin{tabular}{cccc}
    \hline
         model based on&  $T_{h\;avr}$, K& All features& Selected features\\
\hline
         $n_{i}$&  600&  0.972
& 0.977
\\
         &  900&  0.973
& 0.979
\\
         $\eta_{F}$&  600&  0.974
& 0.981\\
         &  900&  0.975
& 0.982\\
    \hline
    \end{tabular}

    \caption{Model performance on the unseen data ($R^{2}$ values on $X_{test}$ dataset).}
    \label{tab:r2_vals}

\end{table}

\begin{figure}[H]
\begin{minipage}[l]{0.32\linewidth}
\includegraphics[width=1\linewidth]{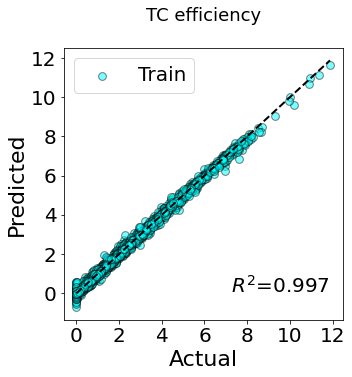} \\ \textbf{a} 
\end{minipage}
\hfill
\begin{minipage}[l]{0.32\linewidth}
\includegraphics[width=1\linewidth]{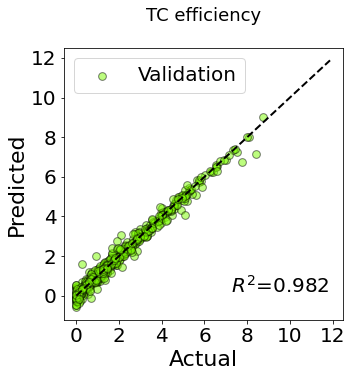} \\ \textbf{b} 
\end{minipage}
\hfill
\begin{minipage}[l]{0.32\linewidth}
\includegraphics[width=1\linewidth]{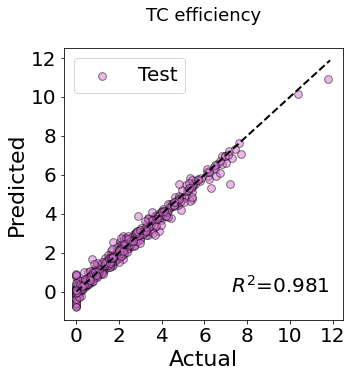} \\ \textbf{c}
\end{minipage}
\caption{Actual VS predicted by ML model values of TC efficiency (in percent) for train, validation and test datasets. Data obtained from model based on Fermi level with physical properties averaged over 300 -- 900 K.}
\label{fig:real_pred}
\end{figure}

Figure \ref{fig:real_pred} indicates how ML predictions of thermocouple efficiency match the actual values. A slight overfitting on train dataset of about \(1.5 \%\) can be seen. Such an overfitting is not crucial. The decrease of the performance on train dataset by the hyperparameters variation led to the simultaneous drop of $R^2$ on both validation and test datasets, resulting in no elimination of the overfitting. Hence, the presented results are the best we were able to observe.
At the same time, model shows the same performance on validation and test datasets pointing to the fact that there is no overfitting on validation dataset.

\subsection{Selected features importance interpretation}

\begin{figure}[H]
\begin{minipage}[l]{0.49\linewidth}
\includegraphics[width=1\linewidth]{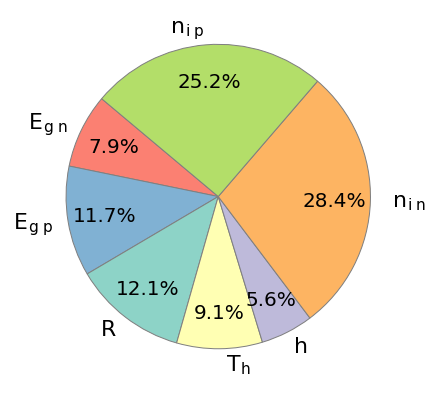} \\ \textbf{a} model based on carriers density
\end{minipage}
\hfill
\begin{minipage}[l]{0.49\linewidth}
\includegraphics[width=1\linewidth]{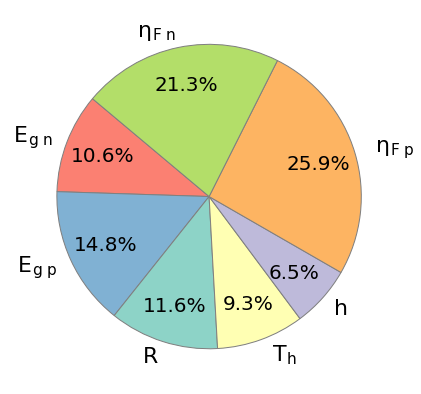} \\ \textbf{b} model based on Fermi level
\end{minipage}
\caption{Averaged importance of selected features. }
\label{fig:pie_charts}
\end{figure}

We averaged the values of  feature importances for models with $T_{h\; avr} = $ 600 and 900 K.  The results are presented in pie charts (Figure \ref{fig:pie_charts}). 
We quantify that around \(1/4\) of $\eta$ is driven by engineering features, while \(3/4\) are due to physical properties. About 50 $\%$ is described by carriers density or Fermi level, with \(n_i\) being a stronger feature than \(\eta_F\) (54 $\%$ against 47 $\%$). 
From \(1/5\) to \(1/4\) of importance corresponds to the energy gap. $E_g$ has an impact of approximately 25 $\%$ for a model based on Fermi level being a slightly more important than $E_g$ in $n_i$--based model where it has 20 $\%$ of importance.

The results point to the importance of the carriers concentration, or the relative Fermi level position, and the energy gap being the physical properties having the largest influence on the final TEG efficiency. The carrier density impacts the electrical conductivity and the electronic contribution to the thermal conductivity, and, via the former, also the Seebeck coefficient, which is generally dependent by the position of the Fermi level in respect of the band edge and hence also on the energy gap. \cite{Lundstrom_S_Eg} Thus, a larger energy gap allows a larger flexibility in tuning of the carrier density and minimizes the effect of the constraint imposed by the intrinsic Fermi level. \cite{JPCC2020, APL2022} These two parameters together have an impact on power factor (PF) and electronic part of thermal conductivity.
It is worth mentioning that none of these properties describe the lattice thermal conductivity. It seems that lattice thermal conductivity plays no significant role in the final efficiency of Half-Heusler-based TEGs.

\begin{figure}[H]
\begin{minipage}[l]{0.4\linewidth}
\includegraphics[width=1\linewidth]{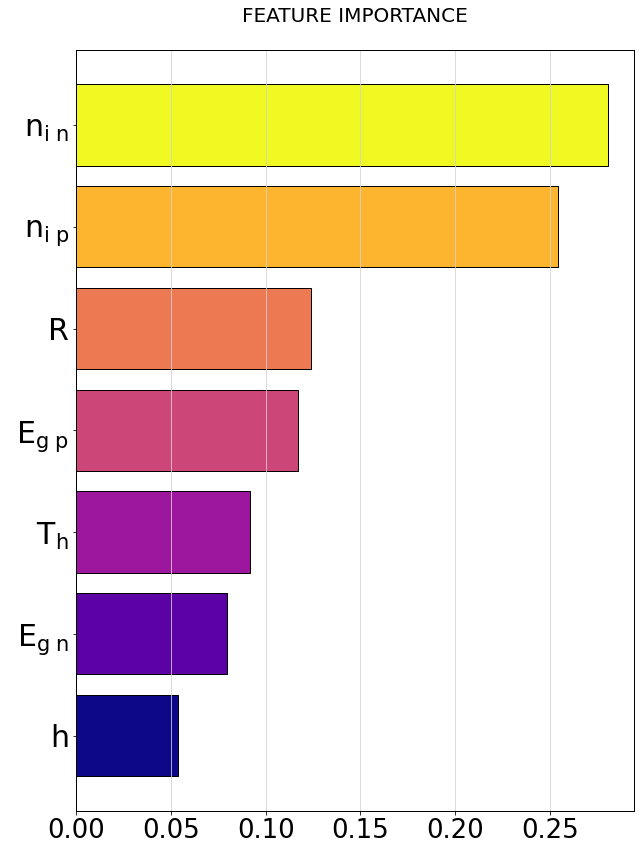} \\ \textbf{a} $T_{h\;avr} = 600 K$
\end{minipage}
\hfill
\begin{minipage}[l]{0.4\linewidth}
\includegraphics[width=1\linewidth]{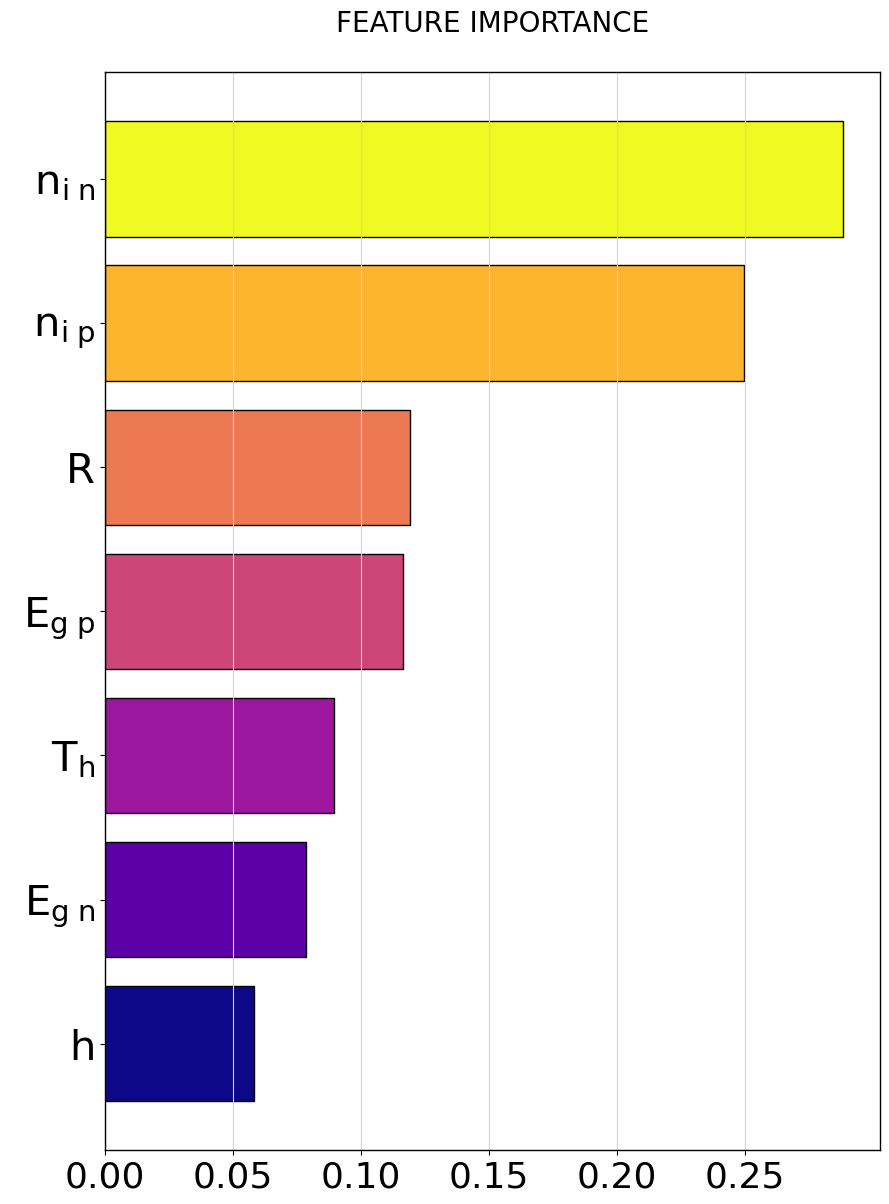} \\ \textbf{b} $T_{h\;avr} = 900 K$
\end{minipage}
\caption{Features importance for a model based on carriers density including only selected features}
\label{fig:feat_imp_n_i_selected}
\end{figure}

\begin{figure}[H]
\begin{minipage}[l]{0.4\linewidth}
\includegraphics[width=1\linewidth]{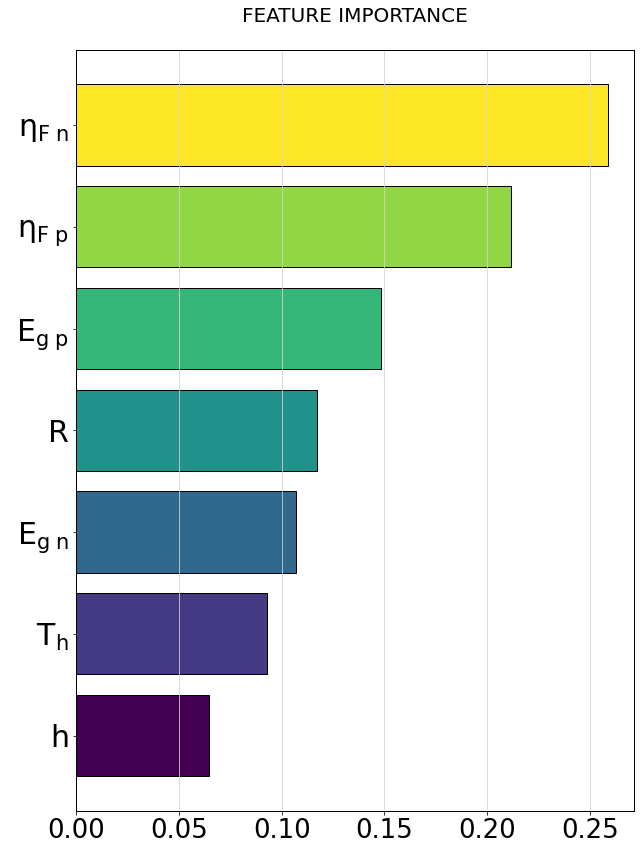} \\ \textbf{a} $T_{h\;avr} = 600 K$
\end{minipage}
\hfill
\begin{minipage}[l]{0.4\linewidth}
\includegraphics[width=1\linewidth]{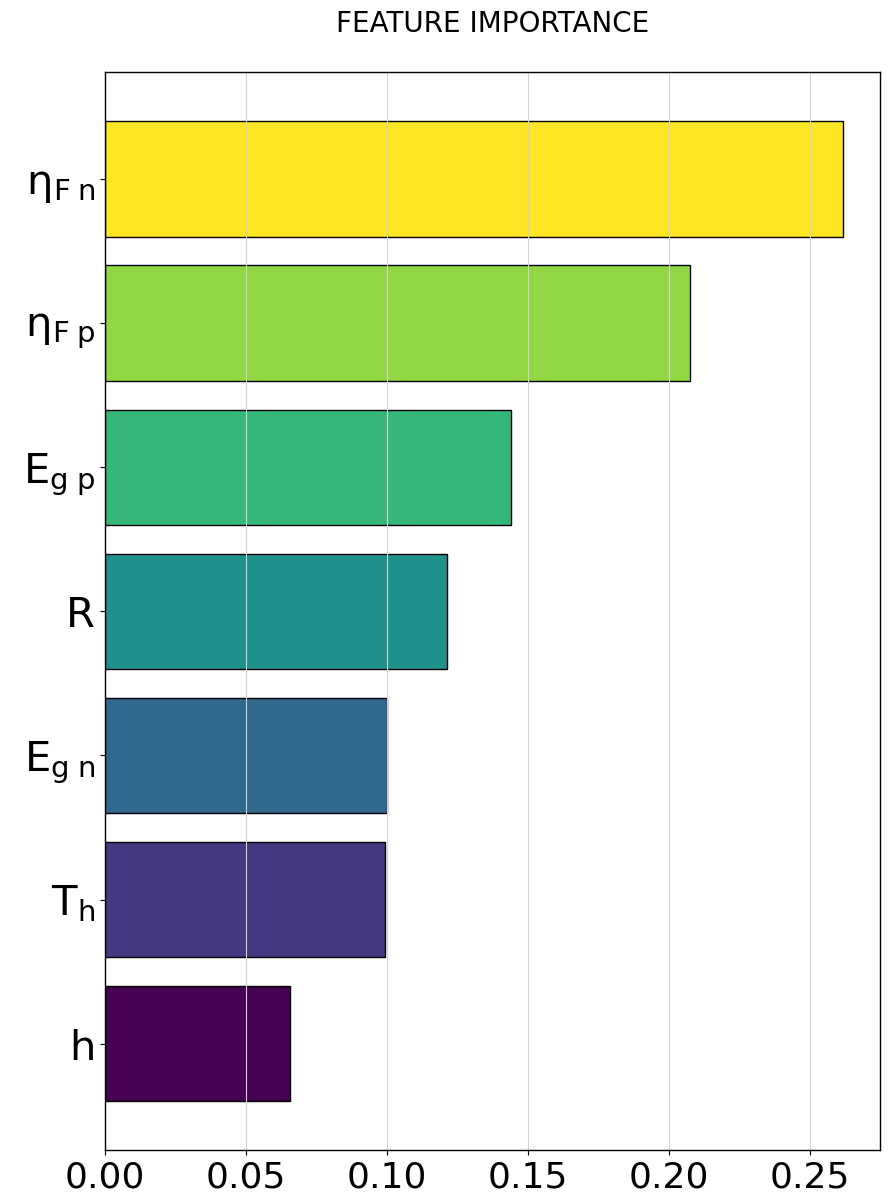} \\ \textbf{b} $T_{h\;avr} = 900 K$
\end{minipage}
\caption{Features importance for a model based on Fermi level including only selected features}
\label{fig:feat_imp_eta_F_selected}
\end{figure}

The importance of selected features is given in Figures \ref{fig:feat_imp_n_i_selected} and \ref{fig:feat_imp_eta_F_selected}.

\subsection{Revealing the optimal values of the selected features via genetic algorithm}

After the most important features have been revealed we did the analysis of $\eta$ dependence on them by collecting all the data points considering the impact of engineering parameters and another physical feature values.

Optimization of physical features has been implemented using a genetic algorithm\cite{GA} (GA) via DEAP\cite{DEAP} library written for Python. GA has been already used\cite{DEMEKE20226633} for optimization of TEG segmented legs composition.
GA is an optimization method inspired by natural selection. It works by iterative evolving a population of candidate solutions to optimize parameters describing the objective function. The process starts with a randomly generated population, each representing a potential solution. Through selection, crossover, and mutation, the algorithm refines the population over multiple generations, seeking to improve performance against a defined objective function. Hyperparameters used for this algorithm tuning are  population size, generations number, mutation rate, and crossover rate.

This approach seeks to find global optimal values for the selected features. These global values are single values that, when applied uniformly across all samples, are expected to yield the highest possible target on average. This approach does not find a unique optimal value for each individual sample but identifies a single set of feature values that maximises the target when applied across the entire dataset. The results are presented in Table\ref{tab:optimized_features}. In the table we perform results for several models, where models 1 and 2 based on $n_i$, 2 and 3 on $\eta_F$. In models 1 and 3 properties were averaged in the range from 300 to 600 K, in models 2 and 4 -- from 300 to 900 K. In the "avr." row we provide the averaged values over all models that can be interpreted as optimal for TEG efficiency maximization.

\begin{figure}[H]
\begin{minipage}[l]{0.49\linewidth}
\includegraphics[width=1\linewidth]{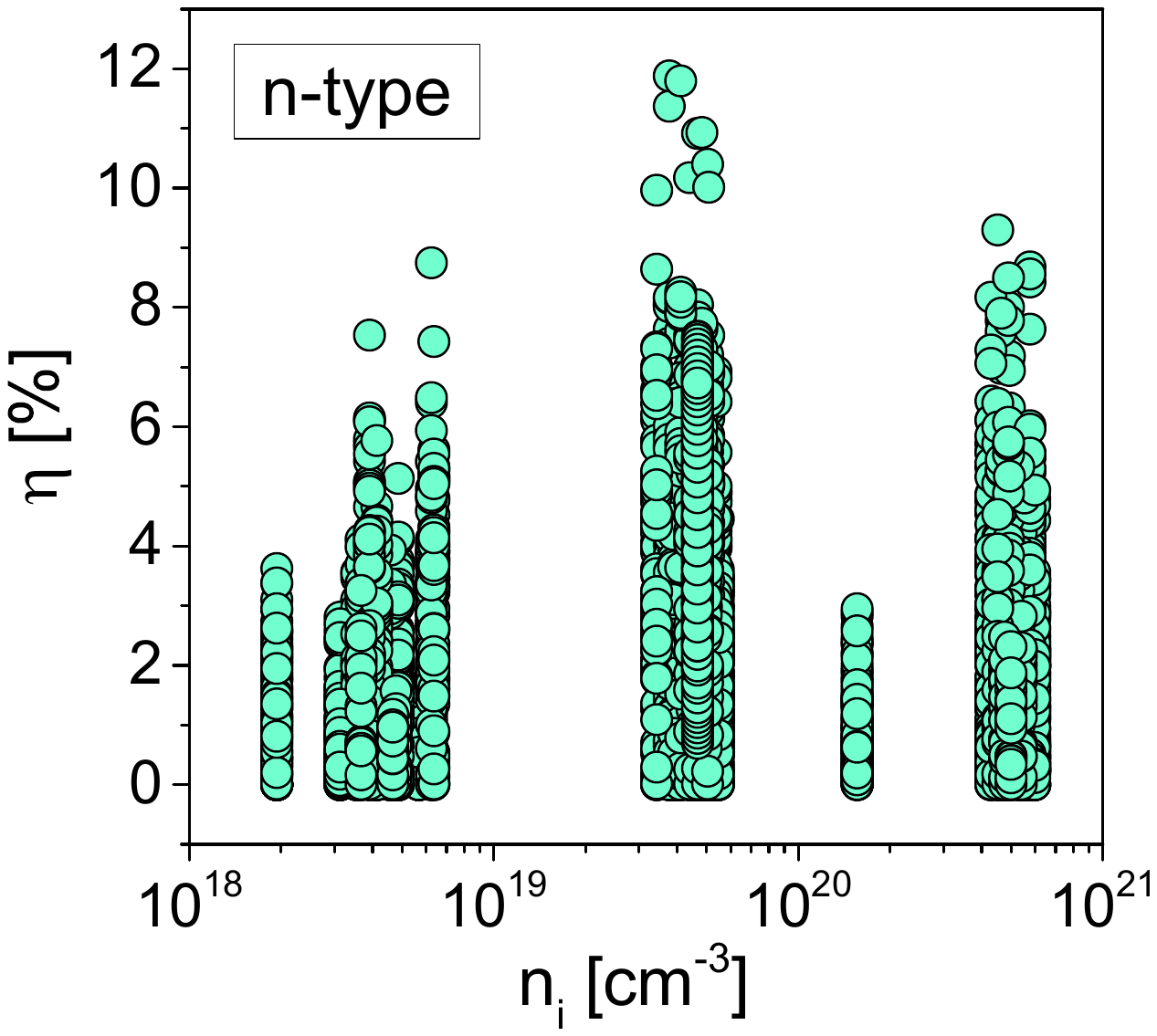} \\ \textbf{a} 
\end{minipage}
\hfill
\begin{minipage}[l]{0.49\linewidth}
\includegraphics[width=1\linewidth]{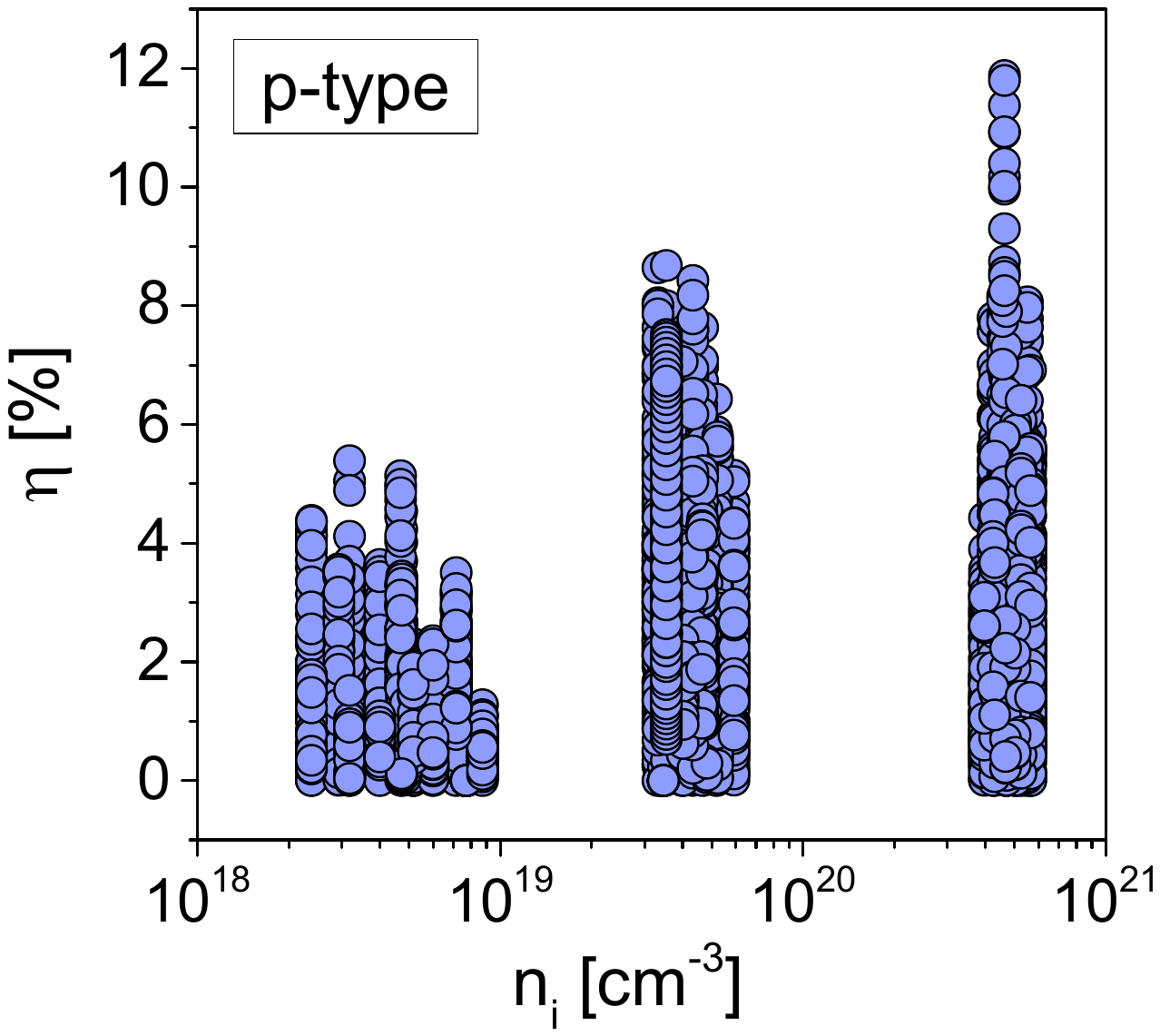} \\ \textbf{b} 
\end{minipage}
\begin{minipage}[l]{0.49\linewidth}
\includegraphics[width=1\linewidth]{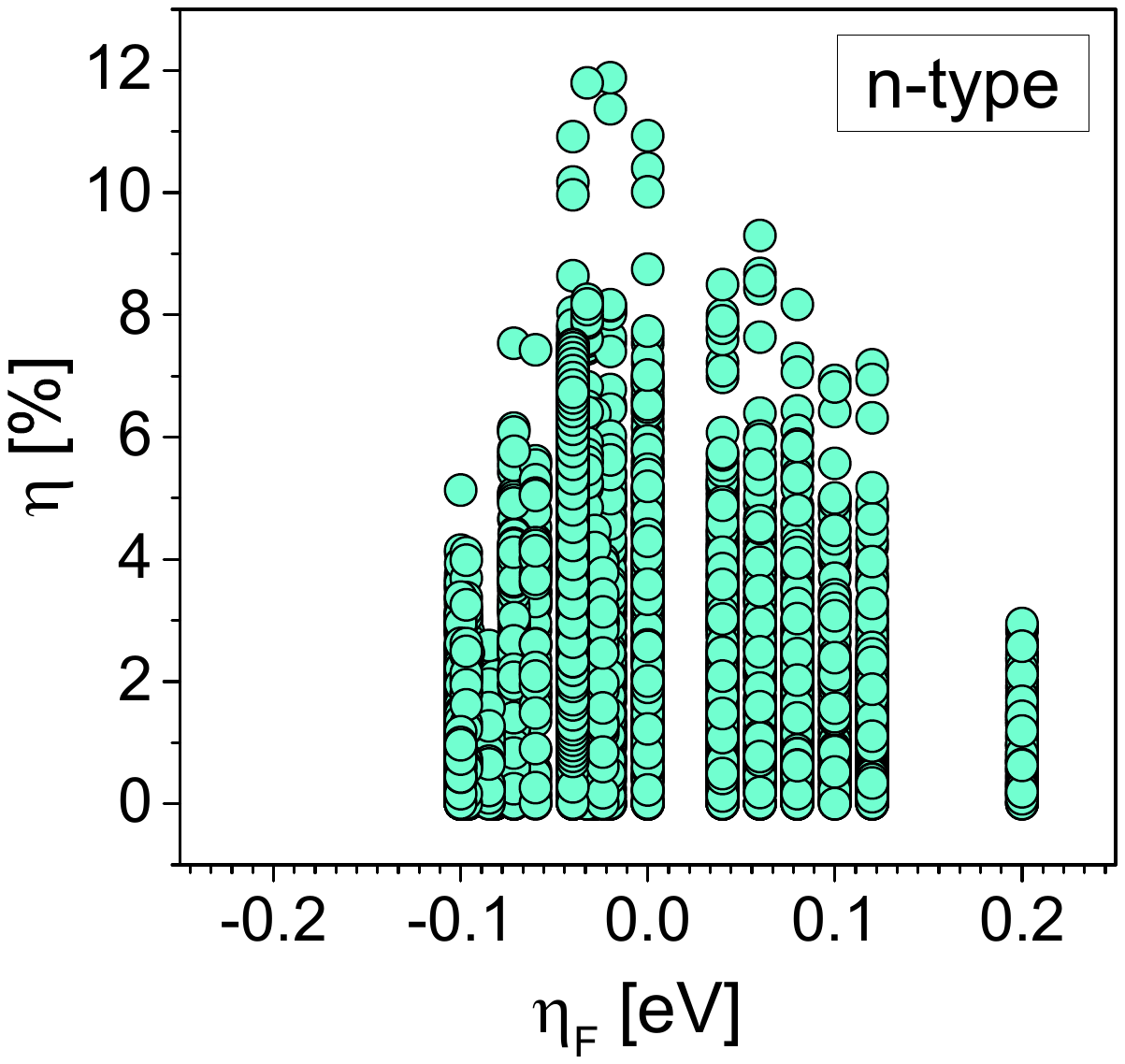} \\ \textbf{c} 
\end{minipage}
\hfill
\begin{minipage}[l]{0.49\linewidth}
\includegraphics[width=1\linewidth]{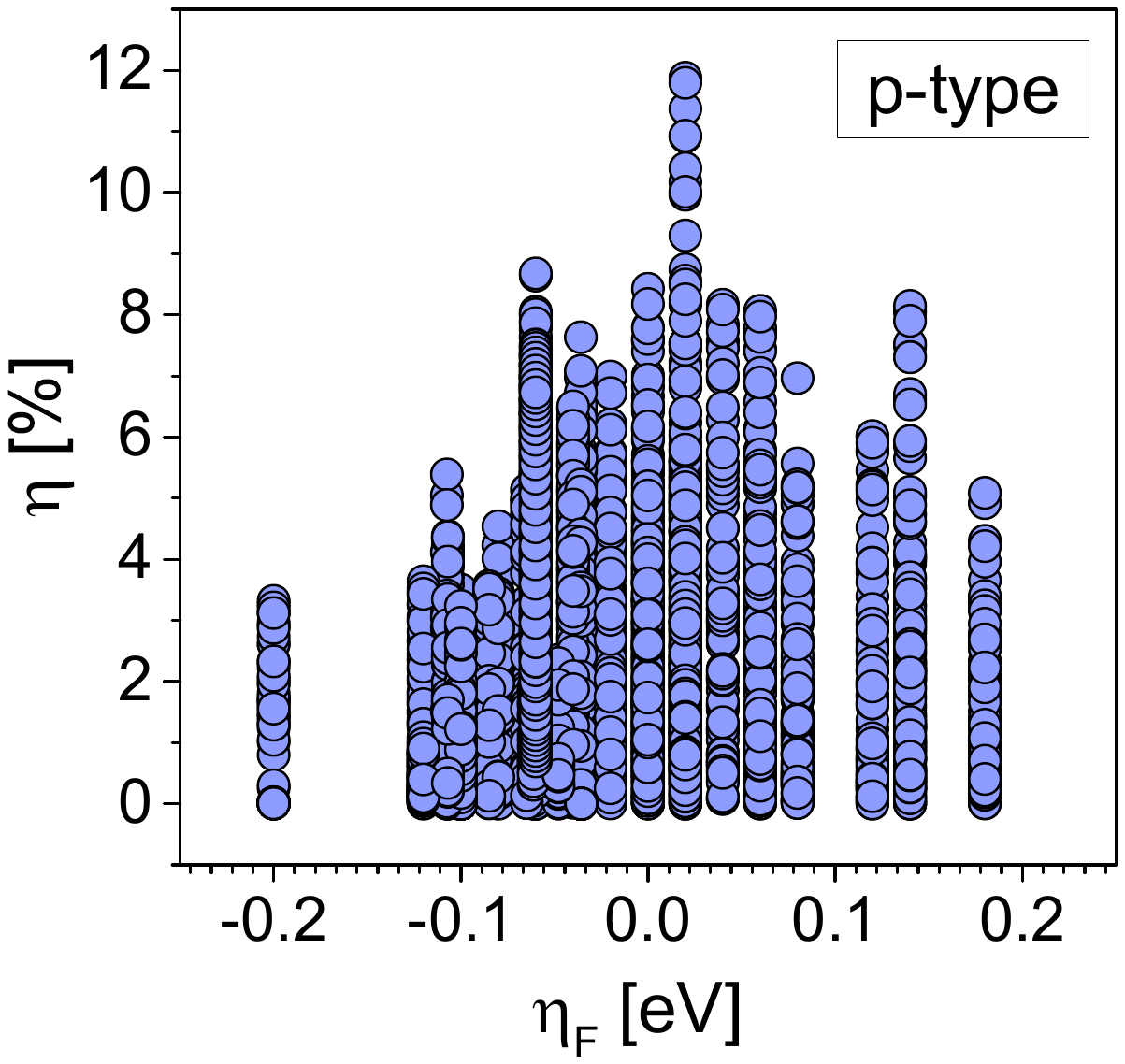} \\ \textbf{d} 
\end{minipage}
\begin{minipage}[l]{0.49\linewidth}
\includegraphics[width=1\linewidth]{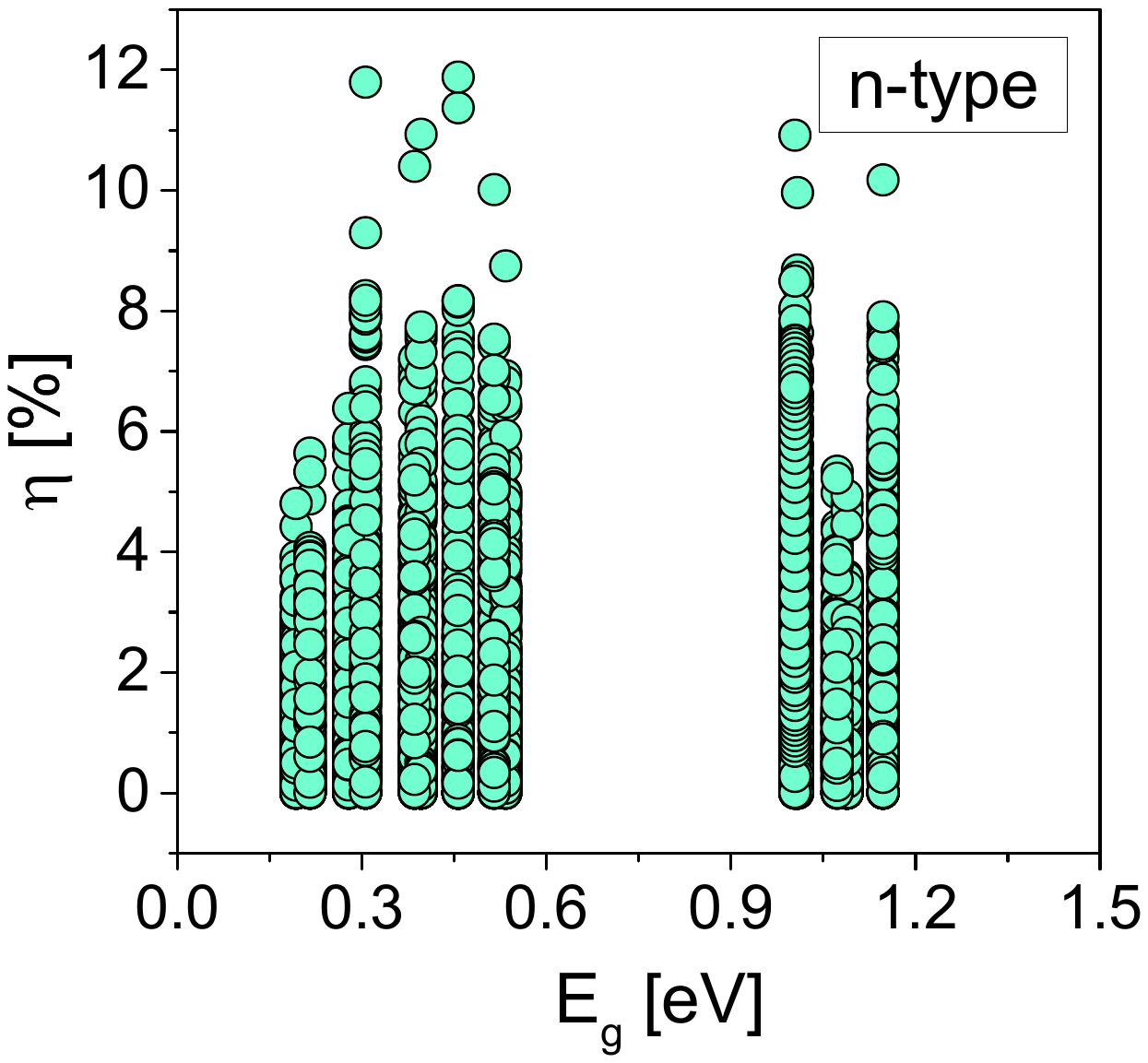} \\ \textbf{e} 
\end{minipage}
\hfill
\begin{minipage}[l]{0.49\linewidth}
\includegraphics[width=1\linewidth]{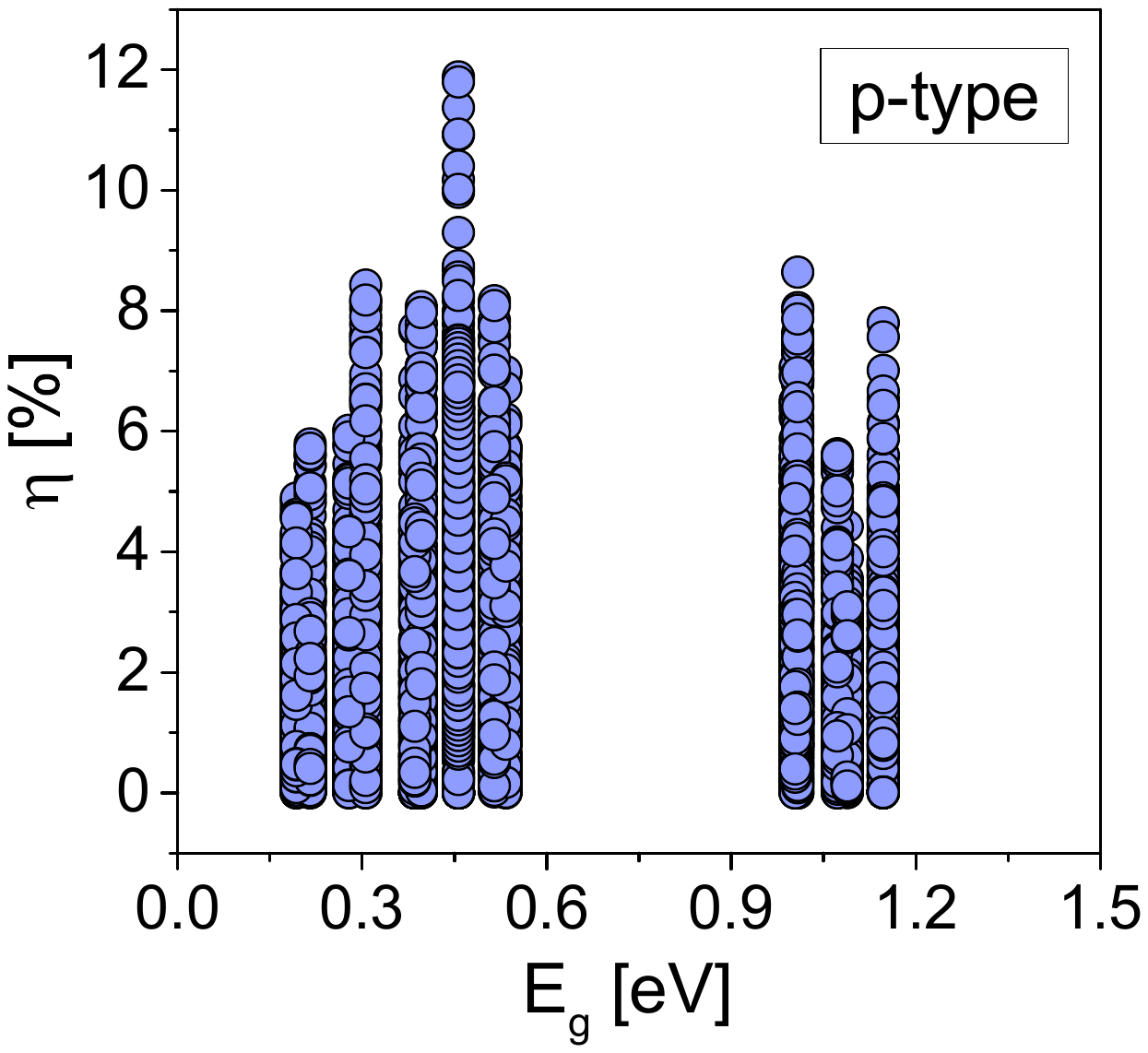} \\ \textbf{f} 
\end{minipage}
\caption{The distribution of thermocouple efficiency values as a dependence of the selected model features: carriers density, Fermi level and energy gap}
\label{fig:eta_dist}
\end{figure}

The distributions of $\eta$ values are given in figure \ref{fig:eta_dist} providing a visualization.

 \begin{table}[H]
\centering
\caption{Features optimised by genetic algorithm  and predicted maximized TEG efficiency. (Units of \(n_i\) are \(cm^{-3}\), \(E_g\) and \(\eta_F\) -- \(eV\), \(R\) -- $Ohm$, \(T_h\) -- \(K\),  \(h\) -- \(mm\), \(\eta_{pred}\) -- \(\%\)).}
\label{tab:optimized_features}
\begin{tabular}{ l  l  l  l  l  l  l  l  l  l  l }
\hline
model& $n_{i\;n}$& $n_{i\;p}$& $\eta_{F\;n}$& $\eta_{F\;p}$& $E_{g\;n}$& $E_{g\;p}$& $R$& $T_{h}$& $h$& $\eta_{pred}$\\
\hline
1 & 4.10e19 & 4.62e20 & - & - & 0.845 & 0.463 & 0.070 & 865 & 4.5 & 10.8 \\
2 & 4.70e19 & 4.63e20 & - & - & 0.907 & 0.466 & 0.158 & 867 & 2.2 & 11.0 \\
3 & - & - & -0.010 & 0.013 & 0.927 & 0.474 & 0.064 & 887 & 4.5 & 12.1 \\
4 & - & - & -0.034 & 0.021 & 0.874 & 0.454 & 0.141 & 867 & 2.5 & 11.1 \\
\hline
avr. & 4.40e19 & 4.63e20 & -0.022 & 0.017 & 0.888 & 0.464 & 0.108 & 871& 3.4 & - \\
\hline

\end{tabular}

\end{table}

 The hyperparameters were selected in order to provide low variation of results within several model runs with specific hyperparameters. We have chosen the following hyperparameters: 
\begin{itemize}
    \item population size -- 100
    \item number of generations -- 200
    \item crossover probability (rate) -- 0.7
    \item mutation probability (rate) -- 0.2
\end{itemize}

We observed in the past that an optimal Fermi level position exists, that maximizes the PF by balancing between electrical conductivity and Seebeck coefficient. \cite{JAP2019} We believe that a similar effect happens here, where the adverse inter-relationship between the transport coefficients calls for a balance between the physical properties, so that an optimization is required rather than a maximization of some parameters. Morever, while our previous finding for optimal material PF called for a Fermi level straight at the edge of the band, with a very few meV window, for both \textit{n-} and \textit{p-}type cases, here we observe a wider window and especially that these window is different for the two polarities. This can be a consequence of the electronic structure details, with nearly parabolic conduction bands and multi-valley complex valence bands. We found that the the carrier density is the most informative physical property, which required to be optimized.

\section{Conclusion}
We propose the data driven approach allowing to reveal the most important physical properties of thermoelectric materials and their optimized values, along with engineering parameters, required for highest TEG efficiency achievement.

The analysis of features available from decision trees feature selection reveiled that the most crucial physical properties for TEGs based on Half-Heuslers are: carriers density or Fermi level. Model based on these parameters along with energy gap, temperature on hot side, leg length and resistance on the external active load achieves the coefficient of determination of 0.98 on test data.

The data generation included DFT, BTE and FEM simulations, whereas the proposed ML tool acts as a direct bridge between DFT and FEM data, eliminating the necessity to use transport coefficients. All necessary information about transport coefficients temperature dependencies, heat loss due to radiation, presence of contacts, full description of temperature  or electric potential fields, etc. is taken into the account by ML model implicitly in the process of learning. Considering high performance we observe, the model shows good ability to address the aforementioned factors without their direct description.

We show that the proposed approach is feasible and convenient. The most time consuming part of the method is data generation. However, we demonstrate that even FEM part, if it is comprehensively optimized, takes approximately 16 hours for 5300 samples generation while running on the  laptop with average technical characteristics. The time required for ML model to make a prediction is at the order of 1 second.
Moreover, the proposed approach can be further improved and expanded. For example, several materials classes can be included in the dataset, as well as the other features can be used as the descriptors. Additional phenomena can be described by the FEM model, e.g. variation of contact materials composition, mechanical stress or chemical diffusion.

The analysis of the features importance gives the room for interpretation. We revealed the most important features to be carriers density or Fermi level, that must be optimized first to obtain a material for effective TEG creation. It can also be interpreted as the indication of the predominant role of electrical conductivity and electronic part of thermal conductivity for Half--Heuslers in the context of energy generation efficiency. Less important features showed their ability to improve the model metrics with energy gap showing itself to be the most useful, however, without a considerable margin. Considering that energy gap has a direct impact on Seebeck coefficient, its utilization in the role of second physical feature makes sense.
The conclusion that lattice thermal conductivity behaves as less important feature points to its minor role in comparison with an electronic part of thermal conductivity. Charge carrier relaxation time or carriers conductivity effective mass can be used to slightly improve the model. It is worth to mention that features selection and interpretation are obtained for a specific material class -- Half-Heuslers. Hence,
it is not a universal conclusion. Whether the scope of our results is general or limited to half-Heuslers-based TEG, requires further research on the path we pave.

Based on the collected data we estimated the optimal values of selected features in a way to achieve the highest TEG efficiency. Genetic algorithm showed itself to be a convenient and fast tool allowing such multiparametric optimization.
In this way, the proposed approach can further contribute to the materials engineering towards the efficient TEGs development.

\begin{acknowledgement}

 P.G. acknowledges funding under the National Recovery and Resilience Plan (NRRP), Mission 04 Component 2 Investment 1.5-NextGenerationEU, Call for tender n. 3277 dated 30/12/2021, Award Number: 0001052 dated 23/06/2022.

\end{acknowledgement}

\begin{suppinfo}

All data and models are available on GitHub: https://github.com/tukmashh/ML-for-thermoelectric-generator-efficiency-prediction

\end{suppinfo}

\bibliography{achemso-demo}

\end{document}